\documentclass[12pt]{article}
\usepackage[dvips]{graphicx}
\usepackage{amssymb}
\usepackage{amsmath}
\usepackage{ulem}
\usepackage[dvipsnames]{xcolor}
\usepackage{soul}
\usepackage{cite}

\makeatletter

\@addtoreset{equation}{section}
\makeatletter

\newcommand{\qed}{\hbox{\rule[-2pt]{6pt}{6pt}}}
\newcommand{\D}{{\rm d}}

\newtheorem{Prop}{Proposition}

\newtheorem{lm}{Lemma}

\setstcolor{red}

\begin{document}

	\begin{titlepage}
		\vfill
		\begin{flushright}
			\today
		\end{flushright}
		
		\vfill
		\begin{center}
			\baselineskip=16pt
			{\Large\bf 
			Junction conditions in scalar-tensor theories\\
			}
			\vskip 0.5cm
			{\large {\sl }}
			\vskip 10.mm
			{\bf  Luis Avil\'es${}^{a, b}$, Hideki Maeda${}^{c}$, and Cristi{\'a}n Mart\'{\i}nez$^{a}$} \\
			
			\vskip 1cm
			{
				${}^a$ Centro de Estudios Cient\'{\i}ficos (CECs), Av. Arturo Prat 514, Valdivia, Chile. \\
				${}^b$ Departamento de F\'{\i}sica, Universidad de Concepci\'on, Casilla 160-C, Concepci\'on, Chile. \\
				${}^c$ Department of Electronics and Information Engineering, Hokkai-Gakuen University, Sapporo 062-8605, Japan.\\
				\texttt{aviles@cecs.cl, h-maeda@hgu.jp, martinez@cecs.cl}
				
			}
			\vspace{6pt}
		\end{center}
		\vskip 0.2in
		\par
		\begin{center}
			{\bf Abstract}
		\end{center}
		\begin{quote}
		We analyze junction conditions at a null or non-null hypersurface $\Sigma$ in a large class of scalar-tensor theories in arbitrary $n(\ge 3)$ dimensions.
After showing that the metric and a scalar field must be continuous at $\Sigma$ as the first junction conditions, we derive the second junctions conditions from the Einstein equations and the equation of motion for the scalar field.
Subsequently, we study $C^1$ regular matching conditions as well as vacuum conditions at $\Sigma$ both in the Jordan and Einstein frames.
Our result suggests that the following configurations may be possible; (i) a vacuum thin-shell at null $\Sigma$ in the Einstein frame, (ii) a vacuum thin-shell at null and non-null $\Sigma$ in the Jordan frame, and (iii) a non-vacuum $C^1$ regular matching at null $\Sigma$ in the Jordan frame.
Lastly, we clarify the relations between the conditions for $C^1$ regularity and also for vacuum $\Sigma$ in the Jordan and Einstein frames.
			\vfill
			\vskip 2.mm
		\end{quote}
	\end{titlepage}

\tableofcontents
\newpage

\section{Introduction}
\label{introduction}

For given two spacetimes, can one attach them at a hypersurface $\Sigma$?
If so, what kind of configurations of $\Sigma$ is possible?
How smooth is the spacetime at $\Sigma$?
These are well-defined problems in gravitation physics and have a variety of applications.
The basic equations to answer these problems are called the {\it junction conditions} which are obtained from the field equations and describe the relation between the discontinuity of the metric and the matter field on the junction hypersurface $\Sigma$ embedded in a bulk spacetime.

In general relativity, a manifestly covariant formalism of the junction conditions has been formulated in the sixties by Israel for non-null (namely, timelike or spacelike) $\Sigma$, which relates the jump of the extrinsic curvature of $\Sigma$ to the energy-momentum tensor for a matter field on $\Sigma$~\cite{Israel:1966rt}.
By the Israel junction conditions, it is shown that the spacetime is $C^1$ (continuously differentiable) and hence regular at $\Sigma$ if and only if there is no matter field on $\Sigma$.
If the spacetime is $C^0$ and hence there is a jump of the extrinsic curvature at $\Sigma$, the matching hypersurface $\Sigma$ is refered to as a {\it thin-shell} or a {\it singular hypersurface}.
In general relativity, a matter field is required on $\Sigma$ for this $C^0$ matching and then $\Sigma$ is refered to as a {\it massive thin-shell}\footnote{In contrast, a {\it vacuum} thin-shell is possible in a class of quadratic curvature gravity called Einstein-Gauss-Bonnet gravity~\cite{vacuumshell}.}.

Obviously, Israel's formulation does not work for null hypersurfaces because the extrinsic curvature is necessarily continuous when $\Sigma$ is null. (See section 3.11.3 in~\cite{Poissonbook}.) 
Indeed, it took more than twenty years until the extension of Israel's formalism for null hypersurfaces was developed by Barrab\`{e}s and Israel \cite{Barrabes:1991ng}.
After being applied in several contexts~\cite{Barrabes:1995nk,Barrabes:1997kh,Barrabes:1998rp,Barrabes:2000es,Barrabes:2000hm,Barrabes:2001vy}, this extension has been reformulated by Poisson~\cite{Poisson:2002nv}. 
Poisson's new formulation makes systematic use of the null generators of the hypersurface and provides a simple characterization of the thin-shell energy-momentum tensor in terms of the jump of the transverse curvature at $\Sigma$.
(See~\cite{j-conditions} for recent developments in the research of junction conditions.)

Alternatively, the junction conditions can also be obtained from the variational principle. 
This method relies on the action principle under Dirichlet boundary conditions for a composite manifold made out of two submanifolds joined at a non-null hypersurface $\Sigma$~\cite{cr1999}. 
The action contains surface terms and its extremum yields not only the field equations in the bulk spacetime but also the junction conditions at $\Sigma$. 
In contrast, derivation of the junction conditions in this method is still unknown in the case where $\Sigma$ is null. 
This is because a general well-defined action principle has not been established on null hypersurfaces.
(See, for instance~\cite{Parattu:2015gga,Lehner:2016vdi,Chakraborty:2016yna}.)

The junction conditions have been studied also in scalar-tensor theories, which are natural generalizations of general relativity and contain a non-minimally coupled scalar field to gravity.
Extensions of Israel's formalism for non-null $\Sigma$ have been presented in a class of scalar-tensor theories~\cite{Sakai:1992ud,Barcelo:2000js,Padilla:2012ze,Nishi:2014bsa}. 
However, these analyses did not consider the case where $\Sigma$ is null. 
Although the junction conditions have been studied both for null and non-null $\Sigma$ in a class of four-dimensional scalar-tensor theories in~\cite{Barrabes:1997kk,Bressange:1997ey}, the analyses were performed only in the Einstein frame and therefore non-minimal couplings for the scalar field were not taken into account. 
As far as the authors know, a study of the junction conditions for null $\Sigma$ in the Jordan frame is absent in the literature in spite of their potential importance for future applications.
One of the purposes of the present paper is to fill this gap.
 
In this article, we study junction conditions at a null or non-null hypersurface $\Sigma$ in a large class of scalar-tensor theories in arbitrary $n(\ge 3)$ dimensions, in which a real scalar field with self-interaction potential is non-minimally coupled to gravity. 
The article is organized as follows. 
In the next section, we will present the action and the field equations of the system both in the Jordan and Einstein frames. 
In Sec.~\ref{sec;non-null}, we will derive the junction conditions in the case where the matching hypersurface $\Sigma$ is non-null and study the $C^1$ regular matching conditions and the vacuum conditions at $\Sigma$ in both frames. 
In Sec.~\ref{sec;null}, we will perform the same analysis as in Sec. \ref{sec;non-null}, but in the case where $\Sigma$ is null. For this purpose, we adopt the formalism presented in \cite{Poisson:2002nv}.
In Sec.~\ref{sec:equivalence}, we will clarify the relations between the conditions for $C^1$ regularity and also for vacuum $\Sigma$ in the Jordan and Einstein frames and apply the result to two different exact solutions.
Our results are summerized in the final section.
Some technical details are presented in two appendices.


\section{Action and field equations in scalar-tensor theories}
\subsection{Preliminaries}
Our basic notations follow~\cite{Poissonbook} and \cite{wald}.
We use the conventions for the curvature tensors such that 
$[\nabla _\rho ,\nabla_\sigma]V^\mu ={R^\mu }_{\nu\rho\sigma}V^\nu$ 
and ${R}_{\mu \nu }={R^\rho }_{\mu \rho \nu }$.
The Minkowski metric has the signature $(-,+,\cdots,+)$ and Greek indices run over all spacetime indices.
We adopt the units such that $c=1$ and $\kappa_n$ denotes the $n$-dimensional gravitational constant.

We consider an $n(\ge 3)$-dimensional Lorentzian (bulk) spacetime ${\cal M}$, of which line element is written as
\begin{align}
\D s_n^2=&g_{\mu\nu}(x)\D x^\mu \D x^\nu.
\end{align}
Let $\partial{\cal M}$ be an $(n-1)$-dimensional non-null hypersurface as a boundary of ${\cal M}$, defined by $\Phi(x)=$constant and let $y^a$ be a set of coordinates on $\partial{\cal M}$.
Since the location of $\partial{\cal M}$ in ${\cal M}$ is described by $x^\mu=x^\mu(y)$, the line element on $\partial{\cal M}$ is given by 
\begin{align}
\D s_{n-1}^2=&h_{ab}(y)\D y^a \D y^b,
\end{align}
where
\begin{align}
h_{ab}(y):=&g_{\mu\nu} e^\mu_a e^\nu_b, \qquad e^\mu_a:=\frac{\partial x^\mu}{\partial y^a}.\label{def-hab-nonnull}
\end{align}
While $g_{\mu\nu}$ and $g^{\mu\nu}$ are respectively used to raise or lower Greek indices, the induced metric $h_{ab}$ and its inverse $h^{ab}$ are used to raise or lower Latin indices, respectively.
For a given vector $v_\mu$, its components on $\partial{\cal M}$ in the coordinates $y^a$ are given by $v_a:=e^\mu_a v_\mu$.
Covariant derivative of $v_a(:=e^\mu_a v_\mu)$ on $\partial{\cal M}$ is given by $D_a v_b\equiv e^\mu_a e^\nu_b (\nabla_\mu v_\nu)$.
\begin{figure}[htbp]
\begin{center}
\includegraphics[width=1.0\linewidth]{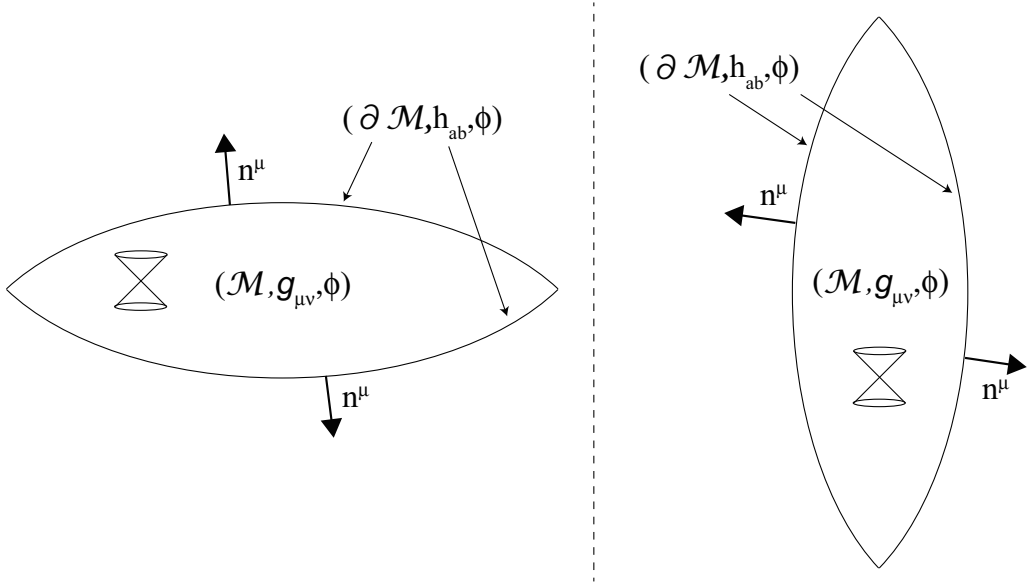}
\caption{\label{Fig-Nonnull-hypersurface} A schematic figure of a spacetime ${\cal M}$ with a spacelike boundary (left) or a timelike boundary (right), denoted by $\partial{\cal M}$.}
\end{center}
\end{figure}

A unit normal vector $n^\mu$ of $\partial{\cal M}$ is given by 
\begin{align} 
n_\mu:=\frac{\varepsilon \nabla_\mu \Phi}{(\varepsilon g^{\rho\sigma}\nabla_\rho \Phi\nabla_\sigma \Phi)^{1/2}}, \label{n-def}
\end{align} 
which satisfies $n^\mu n_\mu=\varepsilon$, where $\varepsilon=1$ ($-1$) corresponds to the case where $\partial{\cal M}$ is a timelike (spacelike) hypersurface. (See Fig.~\ref{Fig-Nonnull-hypersurface}.)
Because $\Phi$ is constant on $\partial{\cal M}$ and hence independent of $y^a$, $n_\mu e^\mu_a=0$ is satisfied.
The Stokes' theorem for a vector field $v^\mu$ in ${\cal M}$ is expressed as
\begin{align} 
\int_{\cal M}\D^{n}x\sqrt{-g}\nabla_\mu v^\mu=\varepsilon \int_{\partial{\cal M}}\D^{n-1}x\sqrt{|h|}n_\mu v^\mu. \label{stokes}
\end{align}

A projection tensor defined by $h_{\mu\nu}:=g_{\mu\nu}-\varepsilon n_\mu n_\nu$ satisfies $h_{\mu\nu}n^\nu=0$ and $h_{ab}=h_{\mu\nu}e^\mu_a e^\nu_b$ (and therefore $h_{\mu\nu}=h_{ab}e^a_\mu e^b_\nu$).
The extrinsic curvature (or the second fundamental form) $K_{\mu\nu}$ of $\partial{\cal M}$ and its trace are defined by 
\begin{align}
K_{\mu\nu}:=&h^{~\rho}_\mu h^{~\sigma}_\nu \nabla_{\rho}n_{\sigma}\left(\equiv\frac12{\cal L}_nh_{\mu\nu}\right), \label{def-exK} \\
K:=&g^{\mu\nu}K_{\mu\nu}=\nabla_\mu n^\mu.
\end{align}

If a symmetric tensor $A_{\mu\nu}$ is tangent to $\partial{\cal M}$, i.e., $A_{\mu\nu}n^\nu\equiv 0$, it admits a decomposition on $\partial{\cal M}$ such that
\begin{align}
A^{\mu\nu}&=A^{ab}e^\mu_a e^\nu_b,
\end{align}
where $A_{ab}(y)=A_{\mu\nu}(x)e^\mu_a e^\nu_b$ is an $(n-1)$-dimensional tensor on $\partial{\cal M}$.
Since $K_{\mu\nu}$ is symmetric and tangent to $\partial{\cal M}$ as $h_{\mu\nu}$, we can write 
\begin{align}
K^{\mu\nu}=K^{ab}e^\mu_a e^\nu_b \, ~~\Leftrightarrow~~\, K_{ab}=K_{\mu\nu}e^\mu_a e^\nu_b, \label{def-Kab-nonnull}
\end{align}
which show $K=g^{\mu\nu}K_{\mu\nu}=h^{ab}K_{ab}$.

\subsection{Jordan frame}
In this work we deal with a class of scalar-tensor theories in $n(\ge 3)$ dimensions characterized by a non-minimally coupled real scalar field $\phi$ endowed with a self-interaction potential $V(\phi)$. 
Our system is described in the Jordan frame by the following action:
\begin{align} 
I_{\rm J}=&\int_{\cal M} \D^{n}x \sqrt{-g}\biggl( f(\phi)R-\frac{1}{2}(\nabla\phi)^2 -V(\phi) \biggl)+\int_{\cal M} \D^{n}x \sqrt{-g}{\cal L}_{\cal M}^{(m)} \nonumber \\
&+2\varepsilon\int_{\partial{\cal M}} \D^{n-1}y \sqrt{|h|}f(\phi)K,\label{J-action}
\end{align}
where $(\nabla\phi)^2:=g^{\mu \nu}(\nabla_\mu \phi)( \nabla_\nu \phi)$ and $\sqrt{-g}{\cal L}_{\cal M}^{(m)}$ is the Lagrangian density for matter fields other than $\phi$. 
The last term in Eq.~(\ref{J-action}) is a boundary term leading a well-defined action principle under Dirichlet boundary conditions, $\delta g_{\mu\nu}|_{\partial{\cal M}}=0=\delta \phi|_{\partial{\cal M}}$. This term will be used to provide an alternative derivation of the junction conditions for non-null hypersurfaces in Section~\ref{otherderivation}.
For simplicity, here we don't consider the case where the boundary $\partial{\cal M}$ consists of several spacelike and timelike portions.

The action (\ref{J-action}) provides the following field equations in the Jordan frame:
\begin{align}
&\! \! \! \! 2f(\phi)G_{\mu\nu}+g_{\mu\nu}\biggl(\frac{1}{2}(\nabla\phi)^2+V(\phi) 
\biggl) \nonumber \\
&~~~~~~~~~~~~~~~-(\nabla_\mu \phi)(\nabla_\nu \phi) -2\nabla_\mu\nabla_\nu 
f(\phi) +2g_{\mu\nu}\Box f(\phi)=T_{\mu\nu},\label{EFE-J}\\
&\Box\phi +f'(\phi)R-V'(\phi)=0,\label{EOM-J}
\end{align}
where a prime denotes derivative with respect to the argument and 
the energy-momentum tensor $T_{\mu\nu}$ for other matter fields is defined by
\begin{align}
T_{\mu\nu}:=-2\frac{\partial {\cal L}_{\cal M}^{(m)}}{\partial 
g^{\mu\nu}}+g_{\mu\nu}{\cal L}_{\cal M}^{(m)}. \label{bulkmatter-T-J}
\end{align}
The action with a typical non-minimally coupled scalar field is realized with the following form of $f(\phi)$:
\begin{equation} 
f(\phi)=\frac{1}{2\kappa_n}-\frac{1}{2}\xi\phi^2, \label{f-typical}
\end{equation}
where $\xi$ is the non-minimal coupling parameter. However, the analysis throughout the text is done for {\it an arbitrary $C^1$ function} $f(\phi)$, namely $f(\phi)$ and its first derivative are both continuous (and hence finite).

In order to simplify the descriptions in the following analysis, we define
\begin{align} 
E_{\mu \nu}:=&2f(\phi)G_{\mu\nu}+g_{\mu\nu}\biggl(\frac{1}{2}(\nabla\phi)^2+V(\phi) \biggl)  \nonumber \\
&-(\nabla_\mu \phi)(\nabla_\nu \phi) -2\nabla_\mu\nabla_\nu f(\phi) +2g_{\mu\nu}\Box f(\phi), \label{EFE}\\
\Pi:=&\Box\phi +f'(\phi)R-V'(\phi), \label{EOM}
\end{align}
so that the field equations (\ref{EFE-J}) and (\ref{EOM-J}) are described as $E_{\mu\nu}=T_{\mu\nu}$ and $\Pi=0$, respectively.

\subsection{Einstein frame}
The scalar-tensor theory in the Jordan frame (\ref{J-action}) is often compared with the following theory:
\begin{align}
I_{\rm E}=&\int_{\cal M} \D^{n}x \sqrt{-{\bar g}}\biggl(\frac{1}{2\kappa_n}{\bar R}-\frac{1}{2}({\bar \nabla}\psi)^2 -{\bar V}(\psi) \biggl)+\int_{\cal M} \D^{n}x \sqrt{-{\bar g}}{\bar {\cal L}}_{\cal M}^{(m)} \nonumber \\
&+\frac{\varepsilon}{\kappa_n}\int_{\partial{\cal M}} \D^{n-1}y \sqrt{|{\bar h}|}{\bar K},\label{action-E}
\end{align}
which is called the Einstein frame of the theory.
As adopted in Eq.~(\ref{action-E}), we will describe geometric quantities in the Einstein frame with bars.

As in the Jordan frame, under an assumption that ${\bar {\cal L}}_{\cal M}^{(m)}$ does not 
depend on $\psi$, the action (\ref{action-E}) provides the following field equations:
\begin{align}
&{\bar G}_{\mu\nu}-\kappa_n\biggl\{({\bar \nabla}_\mu \psi)({\bar 
\nabla}_\nu \psi)-{\bar g}_{\mu\nu}\biggl(\frac{1}{2}({\bar 
\nabla}\psi)^2+{\bar V}(\psi) \biggl)\biggl\}=\kappa_n{\bar 
T}_{\mu\nu},\label{EFE-E}\\
&{\bar \Box}\psi={\bar V}'(\psi),\label{EOM-E}
\end{align}
where ${\bar T}_{\mu\nu}$ is defined by
\begin{align}
{\bar T}_{\mu\nu}:=-2\frac{\partial {\bar {\cal L}}_{\cal 
M}^{(m)}}{\partial {\bar g}^{\mu\nu}}+{\bar g}_{\mu\nu}{\bar {\cal 
L}}_{\cal M}^{(m)}. \label{bulkmatter-T-E}
\end{align}

\subsection{Proper mapping between the Jordan and Einstein frames}
\label{sec:proper}
By a conformal transformation and a redefinition of the scalar field such that 
\begin{align}
{\bar g}_{\mu\nu}=&(2\kappa_n f(\phi))^{2/(n-2)}g_{\mu\nu},\label{conformal-JE}\\
\psi(\phi):=&\pm \int \sqrt{\frac{2(n-1){f'(\phi)}^2+(n-2)f(\phi)}{2(n-2)\kappa_n f(\phi)^2}}\D\phi, \label{def-psi}
\end{align}
the action in the Jordan frame (\ref{J-action}) is mapped to the action (\ref{action-E}), where
\begin{align}
{\bar V}(\psi):=&(2\kappa_n 
f(\phi(\psi)))^{-n/(n-2)}V(\phi(\psi)),\label{Vbar-V-sec2}\\
{\bar {\cal L}}_{\cal M}^{(m)}:=&(2\kappa_n 
f(\phi(\psi)))^{-n/(n-2)}{\cal L}_{\cal M}^{(m)}.\label{TT-M-sec2}
\end{align}
(See Appendix~\ref{App:JtoE} for details.)

Here it should be emphasized that matter fields other than the scalar field may violate a proper mapping between the Jordan and Einstein frames~\cite{fn2007}. 
In general, under the assumption that ${\cal L}_{\cal M}^{(m)}$ is independent of $\phi$, required to give the field equations (\ref{EFE}) and (\ref{EOM}) in the Jordan frame, the conformally transformed action (\ref{action-E}) in the Einstein frame does {\it not} give the equation of motion (\ref{EOM-E}) for $\psi$.
This is because ${\bar {\cal L}}_{\cal M}^{(m)}$ may depend on $\psi$, as seen in Eqs.~(\ref{TT-M-sec2}).
Then, not only ${\bar T}_{\mu\nu}$ depends on $\psi$, but also there appear additional terms in the equation of motion (\ref{EOM-E}) for $\psi$.
An exception is the case where ${\cal L}_{\cal M}^{(m)}$ is for a conformally invariant matter field such as an electromagnetic field in four dimensions.
In such a case, $\sqrt{-g}{\cal L}_{{\cal M}}^{(m)}= \sqrt{-{\bar g}}{\bar {\cal L}}_{{\cal M}}^{(m)}$ holds and then the equation of motion (\ref{EOM-E}) for $\psi$ is obtained in the Einstein frame.

Also, independent of the extra matter fields, a proper mapping between two frames is violated for the following non-minimal coupling
\begin{align}
f(\phi)=-\frac{n-2}{8(n-1)}(\phi-\phi_0)^2, \label{exceptional}
\end{align}
where $\phi_0$ is a constant\footnote{The coupling (\ref{exceptional}) with $\phi_0=0$ makes the sector $\sqrt{-g}\{f(\phi)R-(\nabla\phi)^2/2\}$ in the action (\ref{J-action}) conformal invariant.}.
With this form of $f(\phi)$, the integrand in Eq.~(\ref{def-psi}) is identically zero.
As a result, $\psi$ is constant and there is no inverse transformation $\phi=\phi(\psi)$ even locally.

These observations are summarized in the following lemma, where the assumption (ii) includes the vacuum case, ${\cal L}_{{\cal M}}^{(m)}={\bar {\cal L}}_{{\cal M}}^{(m)}\equiv 0$.
\begin{lm}
\label{lm:correspondence-basic}
Suppose that\\
(i) $f(\phi)$ is a $C^1$ function and not in the exceptional form (\ref{exceptional}), and \\
(ii)  $\sqrt{-g}{\cal L}_{{\cal M}}^{(m)}= \sqrt{-{\bar g}}{\bar {\cal L}}_{{\cal M}}^{(m)}$ holds.\\
Then, there is a proper mapping between the Einstein frame (\ref{action-E}) and the Jordan frame (\ref{J-action}) by a conformal transformation (\ref{conformal-JE}) and redefinitions (\ref{def-psi})--(\ref{TT-M-sec2}).
\end{lm}

\section{Junction conditions for non-null hypersurfaces}
\label{sec;non-null}

\subsection{Setup}
\begin{figure}[htbp]
\begin{center}
\includegraphics[width=0.7\linewidth]{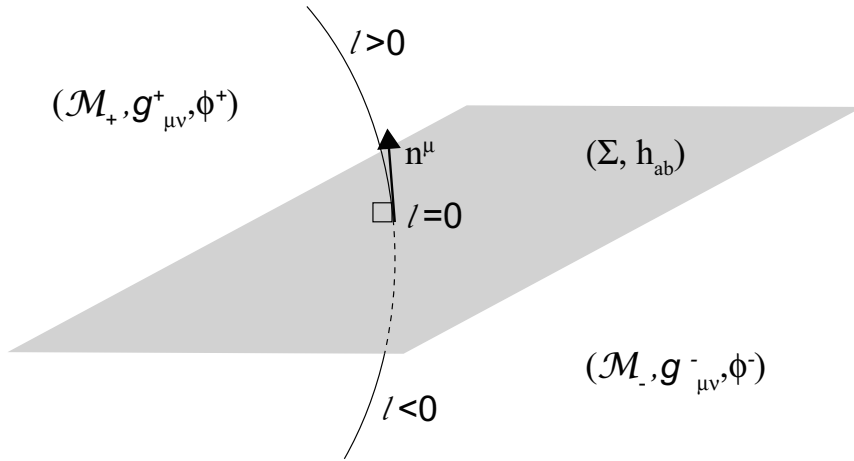}
\caption{\label{Fig-Nonnull-hypersurface2} A non-null hypersurface $\Sigma$ partitions a spacetime into two regions ${\cal M}_+$ and ${\cal M}_-$.}
\end{center}
\end{figure}

We consider a non-null hypersurface $\Sigma$ which partitions a spacetime into two regions ${\cal M}_+$ and ${\cal M}_-$. 
(See~Fig.~\ref{Fig-Nonnull-hypersurface2}.)
Hence $\Sigma$ is a part of  both $\partial{\cal M}_+$ and $\partial{\cal M}_-$.
In ${\cal M}_+$, the metric and the scalar field are $g_{\mu\nu}^+$ and $\phi^+$, respectively, which are functions of the coordinates $x^\mu_{+}$.
In ${\cal M}_-$, the metric and the scalar field are $g_{\mu\nu}^-$ and $\phi^-$, respectively, which are expressed in coordinates $x^\mu_{-}$.
We set the same coordinates $y^a$ on both sides of $\Sigma$, and we choose $n^\mu$, the unit normal to $\Sigma$, to point from ${\cal M}_-$ to ${\cal M}_+$.

Now we assume that continuous {\it canonical coordinates} $x^\mu$, which are different from $x_\pm^\mu$, can be introduced in an open region containing both sides of $\Sigma$.
Actually, the metric and scalar field in ${\cal M}_\pm$ are not described as $g_{\mu\nu}^\pm$ and $\phi^\pm$ in terms of $x^\mu$. 
Nevertheless, hereafter in this section, we keep using the same expressions in the canonical coordinates for simplicity as long as there is no risk of confusion.

Here we use distributions to derive the junction conditions.
The hypersurface $\Sigma$ is considered to be pierced by a congruence of geodesics that intersect it orthogonally. The proper distance (or proper time) along the geodesics is denoted by  $l$, and the  parametrization is adjusted so that $l=0$ when the geodesics cross $\Sigma$. Our convention is that $l$ is negative in ${\cal M}_-$ and positive in ${\cal M}_+$.
Now we introduce the Heaviside distribution $\Theta(l)$, equal to $+1$ if $l>0$, $0$ if $l<0$, and indeterminate if $l=0$.
The distribution $\Theta(l)$ satisfies
\begin{eqnarray}
\Theta(l)^2=\Theta(l),~~~\Theta(l)\Theta(-l)=0,~~~\frac{\D\Theta}{\D l}=\delta(l),
\end{eqnarray}
where $\delta(l)$ is the Dirac distribution, which verifies $\delta(l)=\delta(-l)$. It is important to remark 
that $\Theta(l)\delta(l)$ is not defined as a distribution.
The metric $g_{\mu\nu}$ and the scalar field $\phi$ are expressed in the canonical coordinates $x^\mu$ as
\begin{align}
g_{\mu\nu}=&\Theta(l)g_{\mu\nu}^++\Theta(-l)g_{\mu\nu}^-, \label{g-l2} \\
\phi=&\Theta(l)\phi^++\Theta(-l)\phi^-,\label{scalar-l2} 
\end{align}
which are distribution-valued tensors.

Since the following equations hold along the geodesics,
\begin{align}
\varepsilon \D l^2=g_{\mu\nu}\D x^\mu \D x^\nu,\qquad \varepsilon \frac{\partial l}{\partial x^\mu}\D l=g_{\mu\nu}\D x^\nu~,
\end{align}
where $\varepsilon=1~(-1)$ if $\Sigma$ is timelike (spacelike), a displacement away from $\Sigma$ along one of the geodesics is described by $\D x^\mu=n^\mu \D l$, where $n_\mu$ is given by
\begin{eqnarray}
n_\mu=\varepsilon \partial_\mu l \label{n1}
\end{eqnarray}
and $n^\mu n_\mu=\varepsilon$ holds.
The factor $\varepsilon$ in Eq.~(\ref{n1}) is in order for $n^\mu$ to point from ${\cal M}_-$ to ${\cal M}_+$.
In the canonical coordinates $x^\mu$, the following relations hold:
\begin{equation}
[n^\mu]=[e^\mu_a]=0, \label{n2}
\end{equation}
where $e^\mu_a$ is defined by Eq.~(\ref{def-hab-nonnull}).
Here $[X]$ is defined by
\begin{equation}
[X]:= X^+-X^-,
\end{equation}
where $X^\pm$ are $X$'s evaluated either on the $+$ or $-$ side of $\Sigma$.
The first of Eq.~(\ref{n2}) follows from the relation $\D x^\mu=n^\mu \D l$ and the continuity of both $l$ and $x^\mu$ across $\Sigma$, while the second follows from the fact that the coordinates $y^a$ are the same on both sides of $\Sigma$.

\subsection{Continuity of $g_{\mu\nu}$ and $\phi$: First junction conditions}
The metric $g_{\mu\nu}$ and the scalar field $\phi$ are expressed as Eqs.~(\ref{g-l2}) and (\ref{scalar-l2}) in canonical coordinates $x^\mu$, respectively.
Differentiating them, we obtain
\begin{align}
\partial_\rho g_{\mu\nu}=&\Theta(l)\partial_\rho g_{\mu\nu}^++\Theta(-l)\partial_\rho g_{\mu\nu}^-+\varepsilon \delta(l)[g_{\mu\nu}]n_{\rho}, \label{dg-l2}\\
\partial_\mu \phi=&\Theta(l)\partial_\mu \phi^++\Theta(-l)\partial_\mu \phi^-+\varepsilon \delta(l)[\phi]n_{\rho}.\label{dscalar-l2} 
\end{align}
Thus, to removed the last terms in the right-hand sides which generate terms proportional to $\Theta(l)\delta(l)$ in the Einstein equations (\ref{EFE}) and the equation of motion (\ref{EOM}) for $\phi$, we impose continuity of the metric $g_{\mu\nu}$ and the scalar field $\phi$ across $\Sigma$: 
\begin{equation} \label{firstcondition}
[g_{\mu\nu}]=[\phi]=0.
\end{equation}
This set of conditions is dubbed as the first junction conditions. 
By Eq.~(\ref{def-hab-nonnull}), $[g_{\mu\nu}]=0$ is equivalent to $[h_{ab}]=0$, which means that the induced metric on $\Sigma$ is the same on both sides of $\Sigma$.
The difference of the numbers of equations $[g_{\mu\nu}]=0$ and $[h_{ab}]=0$ is $n$, which coincides with the number of the coordinate conditions $[x^\mu]=0$.

Hereafter, we impose the conditions (\ref{firstcondition}) and the derivatives \eqref{dg-l2} and \eqref{dscalar-l2} then become
\begin{align}
\partial_\rho g_{\mu\nu}=&\Theta(l)\partial_\rho g_{\mu\nu}^++\Theta(-l)\partial_\rho g_{\mu\nu}^-, \label{dg-l3}\\
\partial_\mu \phi=&\Theta(l)\partial_\mu \phi^++\Theta(-l)\partial_\mu \phi^-.\label{dscalar-l3} 
\end{align}
Since the metric and the scalar field are continuous across $\Sigma$ in the canonical coordinates $x^\mu$, the tangential derivatives of the metric and scalar field are also continuous.
Thus, if $\partial_\rho g_{\mu\nu}$ and $\partial_\rho\phi$ are to be discontinuous, the discontinuity must be directed along the normal vector $n^\mu$. Therefore, there must  exist a tensor field $\omega_{\mu\nu}$ and a scalar field $M$  such that 
\begin{equation} \label{disc2}
[\partial_\mu g_{\alpha \beta}]= n_{\mu} \omega_{\alpha \beta}, \hspace{0.5cm} [\partial_\mu \phi]= n_\mu M.
\end{equation}
Namely, $\omega_{\mu\nu}$ and $M$ are defined by 
\begin{equation} \label{disc2-2}
\omega_{\alpha \beta}:=\varepsilon n^\mu[\partial_\mu g_{\alpha \beta}], \hspace{0.5cm} M:=\varepsilon n^\mu[\partial_\mu \phi],
\end{equation}
respectively.

\subsection{Discontinuity of geometric and physical quantities}
From Eqs.~(\ref{g-l2}) and (\ref{dg-l3}), we obtain
\begin{align}
\Gamma^\mu_{\nu\rho}&=\Theta(l){\Gamma^+}^{\mu}_{\nu\rho}+\Theta(-l){\Gamma^-}^{\mu}_{\nu\rho},
\end{align}
where ${\Gamma^\pm}^{\mu}_{\nu\rho}$ is the Christoffel symbols constructed from $g_{\mu\nu}^{\pm}$.
Then, a straightforward calculation with Eqs.~(\ref{n1}) and (\ref{disc2}) reveals
\begin{align}
\partial_\sigma\Gamma^\mu_{\nu\rho}&=\Theta(l)\partial_\sigma{\Gamma^+}^{\mu}_{\nu\rho}+\Theta(-l)\partial_\sigma{\Gamma^-}^{\mu}_{\nu\rho}+\varepsilon\delta(l)[\Gamma^\mu_{\nu\rho}]n_\sigma,
\end{align}
where $[\Gamma^\rho_{\sigma\mu}]$ is given by 
\begin{align}
[\Gamma^\rho_{\sigma\mu}]&=\frac12 (\omega^\rho_{~\sigma}n_{\mu}+\omega^\rho_{~\mu}n_{\sigma}-\omega_{\sigma\mu}n^{\rho}).\label{[Gamma]}
\end{align}
By Eqs.~(\ref{n2}) and (\ref{[Gamma]}), we obtain 
\begin{align}
[\nabla_\mu n_{\nu}]&=-[\Gamma^\sigma_{\mu\nu}]n_{\sigma}\nonumber \\
&=\frac12 (\varepsilon\omega_{\mu\nu}-\omega_{\sigma\mu}n_{\nu}n^{\sigma}-\omega_{\sigma\nu}n_{\mu}n^{\sigma}),
\end{align}
and hence the jump of the extrinsic curvature (\ref{def-exK}) and its trace are given by 
\begin{align}
[K_{\mu\nu}]=&h^{~\rho}_{\mu} h^{~\sigma}_{\nu}[\nabla_\rho n_{\sigma}]  \nonumber \\
=&\frac12(\varepsilon\omega_{\mu\nu}-\omega_{\mu\sigma} n^\sigma n_\nu-\omega_{\nu\rho} n^\rho n_\mu+\varepsilon \omega_{\rho\sigma} n^\rho n^\sigma n_\mu n_\nu), \label{Kmunu-omega}\\
[K]=&g^{\mu\nu}[K_{\mu\nu}]=\frac12(\varepsilon\omega_\mu^{~\mu}-\omega_{\mu\nu}n^{\mu}n^{\nu}).
\end{align}
By Eq.~(\ref{def-Kab-nonnull}), we obtain
\begin{align}
[K_{ab}]=[K_{\mu\nu}]e^\mu_a e^\nu_b=\frac12\varepsilon\omega_{\mu\nu}e^\mu_a e^\nu_b.\label{Kab-def}
\end{align}

On the other hand, using Eq.~(\ref{[Gamma]}), we obtain the Riemann tensor as
\begin{align}
R^\rho_{~\sigma\mu\nu}&=\Theta(l){R^+}^{\rho}_{~\sigma\mu\nu}+\Theta(-l){R^-}^{\rho}_{~\sigma\mu\nu}+\delta(l){\tilde R}^\rho_{~\sigma\mu\nu},\label{riemann}
\end{align}
where the $\delta$-function part of the Riemann tensor is given by 
\begin{align}
{\tilde R}^\rho_{~\sigma\mu\nu}&:=\varepsilon([\Gamma^{\rho}_{\sigma\nu}]n_{\mu}-[\Gamma^{\rho}_{\sigma\mu}]n_{\nu}) \nonumber \\
&=\frac{1}{2}\varepsilon(\omega^\rho_{~~\nu}n_{\sigma}n_{\mu}-\omega^\rho_{~~\mu}n_{\sigma}n_{\nu}-\omega_{\sigma\nu}n^{\rho}n_{\mu}+\omega_{\sigma\mu}n^{\rho}n_{\nu}).\label{A}
\end{align}
Equation~(\ref{A}) shows that the $\delta$-function parts of the Ricci tensor and the Ricci scalar are expressed as
\begin{align}
{\tilde R}_{\sigma\nu}=&{\tilde R}^\mu_{~\sigma\mu\nu}=\frac{1}{2}\varepsilon(\omega_{\nu\mu}n^\mu n_{\sigma}+\omega_{\sigma\mu}n^\mu n_{\nu}-\omega_\mu^{~\mu} n_{\sigma}n_{\nu}-\varepsilon\omega_{\sigma\nu}) \nonumber \\
=&-\varepsilon[K_{\sigma\nu}]-[K]n_\sigma n_\nu,\label{A2} \\
{\tilde R}=&g^{\sigma\nu}{\tilde R}_{\sigma\nu}=-2\varepsilon[K].\label{A3} 
\end{align}
From Eqs.~(\ref{A2}) and (\ref{A3}), the $\delta$-function part of the Einstein tensor ${\tilde G}_{\mu\nu}$ is given as
\begin{align}
{\tilde G}_{\mu\nu}=&{\tilde R}_{\mu\nu}-\frac12g_{\mu\nu}{\tilde R} =-\varepsilon\left([K_{\mu\nu}]-h_{\mu\nu}[K]\right). \label{bar-G}
\end{align}

A $C^1$ regular matching of two spacetimes ${\cal M}_+$ and ${\cal M}_-$ at $\Sigma$ is defined by $[g_{\alpha \beta}]=[\partial_\mu g_{\alpha \beta}]=0$. The following lemma provides several different expressions of a $C^1$ regular matching, among which the condition (i) means that the full Riemann tensor is certainly non-singular at $\Sigma$.
\begin{lm}
\label{lm:regularity-non-null}
If $[g_{\alpha \beta}]=0$ holds, the following six conditions are equivalent: (i) ${\tilde R}^\rho_{~\sigma\mu\nu}=0$, (ii) $[K_{\mu\nu}]=0$ , (iii) $[K_{ab}]=0$, (iv) $\omega_{\mu\nu}=0$, (v) $[\partial_\mu g_{\alpha \beta}]=0$, and (vi) $[\Gamma^\rho_{\sigma\mu}]=0$.
\end{lm}
{\it Proof:}
Equation~(\ref{Kab-def}) shows that the conditions (ii), (iii), and (iv) are equivalent.
By Eqs.~(\ref{disc2}) and (\ref{disc2-2}), the conditions (iv) and (v) are equivalent.
Next we show that the conditions (i) and (ii) are equivalent.
By Eq.~(\ref{Kmunu-omega}), $[K_{\mu\nu}]=0$ implies
\begin{align}
\omega_{\mu\nu}=\varepsilon\omega_{\mu\sigma} n^\sigma n_\nu+\varepsilon\omega_{\nu\rho} n^\rho n_\mu-\omega_{\rho\sigma} n^\rho n^\sigma n_\mu n_\nu. \label{omega-K}
\end{align}
Substituting this into Eq.~(\ref{A}), we obtain ${\tilde R}^\rho_{~\sigma\mu\nu}=0$.
On the other hand, if ${\tilde R}^\rho_{~\sigma\mu\nu}=0$ holds, we have ${\tilde R}_{\sigma\nu}={\tilde R}=0$ and then Eqs.~(\ref{A2}) and (\ref{A3}) show $[K_{\mu\nu}]=0$.
Since we have shown that the conditions (i)--(v) are equivalent, we complete the proof by showing that the conditions (iv) and (vi) are equivalent.
The condition (iv) implies the condition (vi) by Eq.~(\ref{[Gamma]}).
The condition (vi) implies the condition (i) by Eq.~(\ref{A}), which is equivalent to the condition (iv).
\qed

\bigskip

In the following subsections, we will derive the junction conditions from the equation of motion (\ref{EOM-J}) for $\phi$ and the Einstein equations (\ref{EFE-J}).
For this purpose, differentiating Eq.~(\ref{dscalar-l3}), we obtain
\begin{align}
\partial_\mu\partial_\nu \phi=&\Theta(l)\partial_\mu\partial_\nu \phi^++\Theta(-l)\partial_\mu\partial_\nu \phi^-+\varepsilon  \delta(l)M n_\mu n_\nu, \label{ddscalar2} 
\end{align}
where we used Eqs.~(\ref{n1}) and (\ref{disc2}).
From the above expression, we obtain
\begin{align}
\nabla_\mu\nabla_\nu \phi=&\Theta(l)\nabla_\mu\nabla_\nu  \phi^++\Theta(-l)\nabla_\mu\nabla_\nu  \phi^-+\varepsilon  \delta(l)M n_\mu n_\nu, \label{ddscalar2-c} \\
\Box \phi=&\Theta(l)\Box \phi^++\Theta(-l)\Box \phi^-+\delta(l)M. \label{ddscalar2-c-2}
\end{align}
Finally, using the following expression;
\begin{align}
\nabla_\mu\nabla_\nu f(\phi)=&f'(\phi)\nabla_\mu\nabla_\nu \phi+f''(\phi)(\nabla_\mu \phi)(\nabla_\nu \phi),
\end{align}
we obtain
\begin{align}
\nabla_\mu\nabla_\nu f(\phi)=&\Theta(l)\nabla_\mu\nabla_\nu f(\phi^+)+\Theta(-l)\nabla_\mu\nabla_\nu f(\phi^-)+\varepsilon f'(\phi) \delta(l)M n_\mu n_\nu, \label{ddf} \\
\Box f(\phi)=&\Theta(l)\Box f(\phi^+)+\Theta(-l)\Box f(\phi^-)+f'(\phi)\delta(l)M. \label{Box-f}
\end{align}

\subsection{Second junction conditions}
\subsubsection{Equation of motion for a scalar field}
Here we derive the junction condition from the equation of motion (\ref{EOM-J}), namely $\Pi=0$, where $\Pi$ is defined by Eq.~(\ref{EOM}).
Using Eqs.~(\ref{scalar-l2}), (\ref{A3}), and (\ref{ddscalar2-c}), we write down $\Pi$ as
\begin{align} \label{equationKG}
\Pi=\Theta(l)\Pi^++\Theta(-l)\Pi^-+\delta (l){\tilde\Pi},
\end{align}
where the $\delta$-function part ${\tilde\Pi}$ is given by 
\begin{align} 
{\tilde\Pi}:=&M-2\varepsilon f'(\phi)[K]. \label{barPi-def}
\end{align}
The equation of motion (\ref{EOM}) on $\Sigma$ gives ${\tilde\Pi}=0$, namely
\begin{equation} \label{Mreg}
M=2\varepsilon f'(\phi)[K].
\end{equation}
We shall refer to this condition as the junction condition from the equation of motion for a scalar field.
This junction condition is a constraint between the metric and scalar field on $\Sigma$.
For a minimally coupled scalar field, namely $f(\phi)=1/(2\kappa_n)$, this condition is simply $M=0$, which means continuity of $n^\mu \partial_\mu\phi$ at $\Sigma$.

\subsubsection{Einstein equations}
Next let us derive the junction conditions from the Einstein equations (\ref{EFE-J}), namely $E_{\mu\nu}=T_{\mu\nu}$, where $E_{\mu\nu}$ is defined by Eq.~(\ref{EFE}).
Using Eqs.~(\ref{bar-G}), (\ref{ddf}), and (\ref{Box-f}), we write down $E_{\mu\nu}$ as
\begin{align}
E_{\mu\nu}=&\Theta(l){E}^+_{\mu\nu}+\Theta(-l){E}^-_{\mu\nu}+\delta(l){\tilde E}_{\mu\nu},\label{E-decomp}
\end{align}
where the $\delta$-function part ${\tilde E}_{\mu\nu}$ is given by 
\begin{align}
{\tilde E}_{\mu\nu}=-2\varepsilon f(\phi)\left([K_{\mu\nu}]-h_{\mu\nu}[K]\right) +2M f'(\phi) h_{\mu\nu}.
\end{align}
We assume that the bulk matter fields (\ref{bulkmatter-T-J}) do not contain the $\delta$-function part such that
\begin{align}
T_{\mu\nu}=\Theta(l){T}^+_{\mu\nu}+\Theta(-l){T}^-_{\mu\nu},\label{T-decomp}
\end{align}
which means that the bulk matter fields do not contribute to the energy-momentum tensor on $\Sigma$.

By Eqs.~(\ref{E-decomp}) and (\ref{T-decomp}), the Einstein equations $E_{\mu\nu}=T_{\mu\nu}$ on $\Sigma$ give ${\tilde E}_{\mu\nu}=0$, namely
\begin{equation} \label{Ereg}
\varepsilon f(\phi)\left([K_{\mu\nu}]-h_{\mu\nu}[K]\right) =M f'(\phi) h_{\mu\nu}.
\end{equation}
We shall refer to Eq.~(\ref{Ereg}) as the junction conditions from the Einstein equations, which are other constraints between the metric and scalar field on $\Sigma$. 
Under the conditions \eqref{Ereg}, there is no matter field on $\Sigma$ other than $\phi$.
We shall describe this situation as ``$\Sigma$ is vacuum'' throughout this paper.
For a minimally coupled scalar field, namely $f(\phi)=1/(2\kappa_n)$, Eq.~(\ref{Ereg}) reduces to $[K_{\mu\nu}]=h_{\mu\nu}[K]$, which is the same as the general relativistic case.

For embedding configurations of $\Sigma$ with ${\tilde E}_{\mu\nu}\ne 0$, the Einstein equations require an additional matter field on $\Sigma$ for consistency.
Then, $\Sigma$ is no more vacuum and the Einstein equations on $\Sigma$ become
\begin{align}
{\tilde E}_{\mu\nu}=t_{\mu \nu}, \label{Tthin}
\end{align}
where $t_{\mu \nu}$ is the energy-momentum tensor of the matter field on $\Sigma$.

We have now derived the junction conditions (\ref{Mreg}) and (\ref{Tthin}) in the scalar-tensor theories (\ref{J-action}) in the Jordan frame, namely
\begin{align}
&-2\varepsilon f(\phi)\left([K_{\mu\nu}]-h_{\mu\nu}[K]\right) +2M f'(\phi) h_{\mu\nu}=t_{\mu \nu}, \label{j-eq-summary1}\\
&M=2\varepsilon f'(\phi)[K].\label{j-eq-summary2}
\end{align}
Since $h_{\mu\nu}$, $K_{\mu\nu}$, and $t_{\mu \nu}$ are symmetric and tangent to $\Sigma$, we can write Eq.~(\ref{j-eq-summary1})  in terms of intrinsic coordinates $y^a$ on $\Sigma$ such that
\begin{align}
-2\varepsilon f(\phi)\left([K_{ab}]-h_{ab}[K]\right) +2M f'(\phi) h_{ab}=t_{ab}, \label{j-eq-summary1-y}
\end{align}
where $h_{ab}$ and $K_{ab}$ are defined by Eqs.~(\ref{def-hab-nonnull}) and (\ref{def-Kab-nonnull}), respectively, and $t_{ab}:=t_{\mu\nu}e^\mu_a e^\nu_b$.

\subsection{Some notes on junction conditions}
\subsubsection{Derivation by the variational principle} \label{otherderivation}
\begin{figure}[htbp]
\begin{center}
\includegraphics[width=0.5\linewidth]{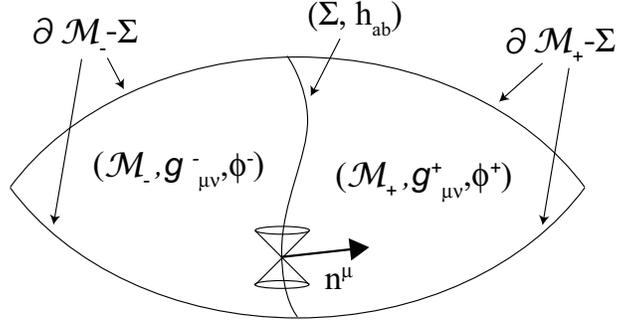}
\caption{\label{Fig-Variation-hypersurface} A schematic figure of a spacetime consisting of two portions ${\cal M_+}$ and ${\cal M_-}$ which are separated by a hypersurface $\Sigma$, of which sides are denoted by $\Sigma_{\pm}$.
This figure shows the case of $\epsilon_+=\epsilon_-=-1$ ($(\partial{\cal M}_{-}-\Sigma_{-})\cup(\partial{\cal M}_{+}-\Sigma_{+})$ is spacelike) and $\varepsilon=1$ ($\Sigma$ is timelike), but there are other possible configurations.}
\end{center}
\end{figure}

Actually, the junction conditions~(\ref{j-eq-summary1}) and (\ref{j-eq-summary2}) can be derived also by the variational principle.
Now the spacetime consists of two parts ${\cal M_+}$ and ${\cal M_-}$ separated by a non-null hypersurface $\Sigma$ such as  Fig.~\ref{Fig-Variation-hypersurface}. In such a spacetime, the action is given by
\begin{align}
I_{\rm J}=&\int_{\cal M_+} \D^{n}x_+ \sqrt{-g^+}\biggl(f(\phi^+)R^+-\frac{1}{2}(\nabla\phi^+)^2 -V(\phi^+)+{\cal L}_{\cal M_+}^{(m)} \biggl) \nonumber \\
&+\int_{\cal M_-} \D^{n}x_- \sqrt{-g^-}\biggl(f(\phi^-)R^--\frac{1}{2}(\nabla\phi^-)^2 -V(\phi^-) +{\cal L}_{\cal M_-}^{(m)}\biggl) \nonumber \\
&+2\epsilon_+ \int_{\partial{\cal M}_+-\Sigma_+} \D^{n-1}z_+ \sqrt{|\zeta^+|}f(\phi^+)K^++2\epsilon_- \int_{\partial{\cal M}_--\Sigma_-} \D^{n-1}z_- \sqrt{|\zeta^-|}f(\phi^-)K^- \nonumber \\
&+2\varepsilon \int_{\Sigma_+} \D^{n-1}y \sqrt{|h|}f(\phi)K^++2\varepsilon \int_{\Sigma_-} \D^{n-1}y \sqrt{|h|}f(\phi)K^- +\int_{\Sigma} \D^{n-1}y \sqrt{|h|}{\cal L}_{\Sigma}^{(m)},\label{action-Sigma-J}
\end{align}
where $\Sigma_{+(-)}$ denotes a side of $\Sigma$ in ${\cal M}_{+(-)}$. (See Appendix~\ref{app:variation-nonnull} for details.)
$\epsilon_+$, $\epsilon_-$, and $\varepsilon$ independently take their values $\pm 1$ and $\phi^\pm|_{\Sigma}=\phi$. 
Here $\sqrt{|h|}{\cal L}_{\Sigma}^{(m)}$ is the Lagrangian density for the matter field other than $\phi$ on $\Sigma$ and we used $z_{\pm}^i$ and $\zeta^{\pm}_{ij}$ for the coordinates and induced metric on the boundary $\partial{\cal M}_{\pm}-\Sigma_{\pm}$, respectively.

Under the assumptions that ${\cal L}_{\cal M_\pm}^{(m)}$ and ${\cal L}_{\Sigma}^{(m)}$ do not depend on $\phi^\pm$ and $\phi$, respectively, variation of the above action, with the boundary condition $\delta g^\pm_{\mu\nu}|_{\partial{\cal M}_{\pm}-\Sigma_{\pm}}=0=\delta \phi^\pm|_{\partial{\cal M}_{\pm}-\Sigma_{\pm}}$, provides the Einstein equations (\ref{EFE-J}) and the equation of motion for the scalar field (\ref{EOM-J}) in the bulk spacetimes ${\cal M_+}$ and ${\cal M_-}$, as well as, the junction conditions~(\ref{j-eq-summary1}) and (\ref{j-eq-summary2}) on $\Sigma$, where $t_{\mu\nu}=t_{ab}e^a_\mu e^b_\mu$ is given by the energy-momentum tensor $t_{ab}$ for other matter fields on $\Sigma$ defined by
\begin{align} \label{otherTs}
t_{ab}:=-2\frac{\partial {\cal L}_{\Sigma}^{(m)}}{\partial h^{ab}}+h_{ab}{\cal L}_{\Sigma}^{(m)}.
\end{align}
The details of derivation are presented in Appendix~\ref{app:variation-nonnull}.

\subsubsection{Comments on the matter field on $\Sigma$}
\label{Sec:matteronSigma}

Here we should comment on the energy-momentum tensor $t_{ab}$ on $\Sigma$.
We have seen in Eq.~\eqref{otherTs} that $t_{ab}$ is obtained from its Lagrangian density in the variational approach.
In such a case, $D^bt_{ab}=0$ holds and hence the energy-momentum conservation is satisfied on $\Sigma$ under the assumptions in the following lemma\footnote{In contrast, the energy-momentum tensor introduced in \eqref{Tthin} has no such a requirement.}. (See Appendix E in~\cite{wald}.)
\begin{lm}
\label{lm:conservation}
Let $\sqrt{|h|}{\cal L}_\Sigma^{(m)}$ be a matter Lagrangian density for a matter field $\Psi$ (not necessary to be a scalar field) on a non-null hypersurface $\Sigma$.
If the bulk action does not contain $\Psi$, then the energy-momentum tensor $t_{ab}$ defined by Eq.~(\ref{otherTs}) satisfies $D^at_{ab}=0$.
\end{lm}
{\it Proof:}
The matter action is given by 
\begin{align} \label{set-theory}
I^{(m)}_\Sigma:=\int_\Sigma \D^{n-1}y \sqrt{|h|}{\cal L}_\Sigma^{(m)}.
\end{align}
The variation of the action~(\ref{set-theory}) on a non-null hypersurface on $\Sigma$ results in the following form:
\begin{align} \label{set-theory1}
\delta I^{(m)}_\Sigma=\int_\Sigma \D^{n-1}y \sqrt{|h|}\left(-\frac12t_{ab}\delta h^{ab}+{\tilde{\cal E}}_{(\Psi)}\delta \Psi\right)+\int_{\partial\Sigma}\D^{n-2}z \sqrt{|{\tilde h}|}\left({\tilde{\cal F}}_{ab}\delta h^{ab}+{\tilde{\cal F}}_{(\Psi)}\delta \Psi\right),
\end{align}
where $z^i$ and $|{\tilde h}|$ are the coordinates and the determinant of the induced metric at the boundary of $\Sigma$, respectively.
By assumptions, variation of the bulk action does not generate any term proportional to $\delta \Psi$ on $\Sigma$.
Thus, the action principle on $\Sigma$ with the boundary conditions $\delta h^{ab}=\delta \Psi=0$ at $\partial\Sigma$ gives ${\tilde{\cal E}}_{(\Psi)}=0$ as an equation of motion for $\Psi$ on $\Sigma$.

Using the equation of motion ${\tilde{\cal E}}_{(\Psi)}=0$ and the boundary conditions $\delta h^{ab}=\delta \Psi=0$ at $\partial\Sigma$, we can rewrite the variation (\ref{set-theory1}) as
\begin{align} \label{set-theory2}
\delta I^{(m)}_\Sigma=-\frac12\int_\Sigma \D^{n-1}y \sqrt{|h|}t_{ab}\delta h^{ab}.
\end{align}
Now we use the fact that the action is diffeomorphism invariant on $\Sigma$, namely the coordinate invariant, and therefore $\delta I^{(m)}_\Sigma=0$ holds for such variations.
If the differomorphism is generated by an infinitesimal vector field $w^a$ on $\Sigma$, we have $\delta h^{ab}={\cal L}_wh^{ab}=2D^{(a}w^{b)}$, where ${\cal L}_w$ is the Lie derivative along $w^a$.
Then, from Eq.~(\ref{set-theory2}), $\delta I^{(m)}_\Sigma=0$ implies
\begin{align} 
0=&-\frac12\int_\Sigma \D^{n-1}y \sqrt{|h|}t_{ab}D^{(a}w^{b)}=-\int_\Sigma \D^{n-1}y \sqrt{|h|}t_{ab}D^{a}w^{b} \nonumber \\
=&-\int_\Sigma \D^{n-1}y \sqrt{|h|}\left(D^{a}(t_{ab}w^{b})-(D^{a}t_{ab})w^{b}\right) \nonumber \\
=&-\varepsilon\int_{\partial\Sigma} \D^{n-2}z \sqrt{|{\tilde h}|}n^{a}t_{ab}w^{b}+\int_\Sigma \D^{n-1}y \sqrt{|h|}(D^{a}t_{ab})w^{b} \nonumber \\
=&\int_\Sigma \D^{n-1}y \sqrt{|h|}(D^{a}t_{ab})w^{b},
\end{align}
where we used the Stokes' theorem (\ref{stokes}) and the boundary condition $w^{a}=0$ at $\partial\Sigma$.
Since the above equation is satisfied for an arbitrary generator $w^a$, $D^{a}t_{ab}=0$ is concluded.
\qed

The junction conditions from the Einstein equations (\ref{Tthin}) can be written as ${\tilde E}_{ab}=t_{ab}$, where
\begin{align}
{\tilde E}_{ab}:=&{\tilde E}_{\mu\nu}e^\mu_a e^\nu_b \nonumber \\
=&-2\varepsilon f(\phi)\left([K_{ab}]-h_{ab}[K]\right) +2M f'(\phi) h_{ab}.
\end{align}
Divergence of ${\tilde E}_{ab}$ is written as
\begin{align}
D^a{\tilde E}_{ab}=-2\varepsilon f(\phi)[{R}_{\nu\sigma}]e^\nu_{~b}n^\sigma-2\varepsilon f'(\phi)(D^a\phi)\left([K_{ab}]-h_{ab}[K]\right)+2D_b(M f'(\phi)), \label{j-eq-summary1-y2}
\end{align}
where we used the Codazzi equation:
\begin{align}
{R}_{\mu\nu\rho\sigma}e^\mu_{~a}e^\nu_{~b}e^\rho_{~c}n^\sigma=D_{a}K_{bc}-D_{b}K_{ac}~~\Rightarrow~~{R}_{\nu\sigma}e^\nu_{~b}n^\sigma=D^cK_{bc}-D_{b}K.\label{gc2}
\end{align}
We note that, if $[\partial_\rho g_{\mu\nu}]$ is non-vanishing at $\Sigma$, we have $D^a{\tilde E}_{ab}\ne 0$ in general (even in general relativity).
Therefore, by Lemma~\ref{lm:conservation}, the junction conditions ${\tilde E}_{ab}=t_{ab}$ require (i) an embedding configuration satisfying $D^a{\tilde E}_{ab}\equiv 0$ or (ii) violation of the assumption in Lemma~\ref{lm:conservation}, which means that ${\cal L}_\Sigma^{(m)}$ depends on $\phi$.

While to achieve the case (i) is rather difficult in the Jordan frame, there is a simple example of such configurations of $\Sigma$ in the Einstein frame.
In the Einstein frame, where $f(\phi)=1/2\kappa_n$ holds, Eq.~(\ref{j-eq-summary1-y2}) reduces to
\begin{align}
D^a{\tilde E}_{ab}=-\frac{\varepsilon}{\kappa_n}[{R}_{\nu\sigma}]e^\nu_{~b}n^\sigma. \label{DaEab-3-2-Ein}
\end{align}
and hence $[R_{\nu\sigma}]e^\nu_{b} n^\sigma=0$ is required for $D^a{\tilde E}_{ab}=D^at_{ab}=0$.
This condition is accomplished for any non-null $\Sigma$ embedded in an Einstein space because $R_{\mu\nu}\propto g_{\mu\nu}$ implies $[R_{\nu\sigma}]e^\nu_{b} n^\sigma\propto [g_{\nu\sigma}]e^\nu_{b} n^\sigma=0$.
The condition $[R_{\nu\sigma}]e^\nu_{b} n^\sigma=0$ is also satisfied for the following spacetime:
\begin{align}
\D s^2=-F(r)\D t^2+F(r)^{-1}\D r^2+r^2\gamma_{ij}\D z^i\D z^j,
\end{align}
if the shell is described by $t=t(\tau)$ and $r=r(\tau)$, where $\tau$ is a parameter and $\gamma_{ij}\D z^i\D z^j$ is the line element on an $(n-2)$-dimensional Einstein space.

Next let us consider the case (ii), namely the case where ${\cal L}_\Sigma^{(m)}$ depends on $\phi(=\phi^{\pm}|_\Sigma)$.
As an example, we consider ${\cal L}_{\Sigma}^{(\phi)}(\in{\cal L}_{\Sigma}^{(m)})$ for $\phi$ on $\Sigma$ as in the bulk:
\begin{align}
{\cal L}_{\Sigma}^{(\phi)}:=&f(\phi){\cal R}-\frac{1}{2}(D\phi)^2 -V(\phi),
\end{align}
where $(D\phi)^2:=h^{ab}(D_a\phi)(D_b\phi)$ and ${\cal R}$ is the Ricci scalar constructed from $h_{ab}$.
Then, the equation of motion for $\phi$ on $\Sigma$ becomes 
\begin{align} \label{equationKGnonnull-matter}
D_aD^a\phi+f'(\phi){\cal R}-V'(\phi)=-{\tilde\Pi},
\end{align}
where ${\tilde\Pi}$ is defined by Eq.~(\ref{barPi-def}).
In this case, the energy-momentum tensor $t_{ab}^{(\phi)}$ for $\phi$ on $\Sigma$, defined by 
\begin{align}
t^{(\phi)}_{ab}:=-2\frac{\partial {\cal L}_{\Sigma}^{(\phi)}}{\partial h^{ab}}+h_{ab}{\cal L}_{\Sigma}^{(\phi)},
\end{align}
satisfies $D^at_{ab}^{(\phi)}\ne 0$ if  ${\tilde\Pi}\neq 0$.
However in general, it is highly nontrivial whether there exists a configuration of $\Sigma$ with this $t_{ab}$.
There is even a possibility that the junction conditions ${\tilde E}_{ab}=t_{ab}$ and the equation of motion (\ref{equationKGnonnull-matter}) do not allow any solution.

In summary, when the energy-momentum tensor of a matter field on $\Sigma$ is assumed to come from a Lagrangian density, the junction conditions (\ref{Tthin}) in the Jordan frame severely constrain the configuration of $\Sigma$ with non-vanishing $[\partial_\rho g_{\mu\nu}]$.

\subsection{Conditions for $C^1$ matching and vacuum $\Sigma$}

\subsubsection{Jordan frame}

Now let us study the conditions for a $C^1$ matching and also for vacuum $\Sigma$.
The following proposition shows that $M=t_{ab}=0$ is a necessary and sufficient condition for a $C^1$ matching at $\Sigma$ in the Jordan frame if $f(\phi)\ne 0$ holds there.
\begin{Prop}
\label{Prop:C1matching}
({\it J-regularity at non-null $\Sigma$.})
Suppose in the Jordan frame that \\
(i) $f(\phi)$ is a $C^1$ function,\\ 
(ii) $[g_{\mu\nu}]=[\phi]=0$ holds at a non-null hypersurface $\Sigma$, and\\
(iii) the second junction conditions at $\Sigma$ are given by Eqs.~(\ref{j-eq-summary2}) and (\ref{j-eq-summary1-y}).\\
Then, the $C^1$ regularity at $\Sigma$ implies $M=t_{ab}=0$. Moreover, 
$M=t_{ab}=0$ and $f(\phi)\ne 0$ at $\Sigma$ implies the $C^1$ regularity at $\Sigma$.
\end{Prop}
{\it Proof}.
By Lemma~\ref{lm:regularity-non-null}, the $C^1$ regularity at $\Sigma$ is equivalent to $[K_{ab}]=0$.
Then, the proposition follows from Eqs.~(\ref{j-eq-summary2}) and (\ref{j-eq-summary1-y}).
\qed

In the special case where $f(\phi)= 0$ holds at $\Sigma$, $M=t_{ab}=0$ is just a necessary condition for a $C^1$ regular matching.
Actually, $M=t_{ab}=0$ only implies $f'(\phi)[K]=0$ in this case.

The following proposition shows the conditions for vacuum $\Sigma$ ($t_{ab}\equiv 0$).

\newpage

\begin{Prop}
\label{Prop:C1-non-null-J}
({\it J-vacuum at non-null $\Sigma$.})
Let $\phi_\Sigma$ be the value of $\phi$ at a non-null hypersurface $\Sigma$.
Then, under the assumptions (i)--(iii) in Proposition~\ref{Prop:C1matching}, $t_{\mu \nu}\equiv 0$ is realized at 
$\Sigma$ only in one of the following three cases:\\
(I) $[K_{\mu\nu}]=M=0$,  \\
(II) $f'(\phi_\Sigma)=f(\phi_\Sigma)=M=0$, or \\
(III) $2(n-1){f'(\phi_\Sigma)}^2+(n-2)f(\phi_\Sigma)=0$, 
$(n-1)[K_{\mu\nu}]=[K] h_{\mu\nu}$, and $M=2\varepsilon f'(\phi_\Sigma)[K]$.
\end{Prop}
{\it Proof}.
With $t_{\mu \nu}\equiv 0$, the junction conditions 
(\ref{j-eq-summary1}) and (\ref{j-eq-summary2}) show
\begin{align}
&-f(\phi)\left([K_{\mu\nu}]-h_{\mu\nu}[K]\right) +2f'(\phi)^2[K] 
h_{\mu\nu}=0,\label{j-eq-summary1-C1}
\end{align}
of which trace gives
\begin{align}
\left\{(n-2)f(\phi) +2(n-1)f'(\phi)^2\right\}[K]=0
\end{align}
and hence $[K]=0$ or $2(n-1){f'(\phi_\Sigma)}^2+(n-2)f(\phi_\Sigma)=0$ 
is required.

If $[K]=0$, then $[K_{\mu\nu}]=0$ and $M=0$ are concluded by 
Eqs.~(\ref{j-eq-summary2}) and (\ref{j-eq-summary1-C1}), since $f(\phi)$ 
is in the $C^1$ class and hence both $f(\phi_\Sigma)$ and 
$f'(\phi_\Sigma)$ are finite.

If $2(n-1){f'(\phi_\Sigma)}^2+(n-2)f(\phi_\Sigma)=0$, the junction 
conditions (\ref{j-eq-summary1}) and (\ref{j-eq-summary2}) reduce to
\begin{align}
{f'(\phi_\Sigma)}^2\left\{(n-1)[K_{\mu\nu}]-[K] h_{\mu\nu}\right\}&=0,\\
M-2\varepsilon f'(\phi_\Sigma)[K]&=0,
\end{align}
and hence there are two possibilities 
$f'(\phi_\Sigma)=f(\phi_\Sigma)=M=0$ or $(n-1)[K_{\mu\nu}]=[K] 
h_{\mu\nu}$ with $M=2\varepsilon f'(\phi_\Sigma)[K]$.
\qed

While the case (I) in Proposition~\ref{Prop:C1-non-null-J} is the same as 
that in the Einstein frame, the cases (II) and (III) are characteristic 
in the Jordan frame, which suggest the possibility of a vacuum thin-shell, where the spacetime is vacuum but $C^0$ at $\Sigma$.
(See~\cite{vacuumshell} for such a vacuum thin-shell in Einstein-Gauss-Bonnet gravity.)
While the constraint on the jump of the extrinsic curvature at $\Sigma$ is different from $[K_{\mu\nu}]=0$ in the case (III), there is no constraint  $[K_{\mu\nu}]$ in the case (II).
We note that the first condition in the case (III) is always satisfied in the theory with the non-minimal coupling (\ref{exceptional}), which does not admit the Einstein frame.

\subsubsection{Einstein frame} 
While the total action in the Jordan frame is given by Eq.~(\ref{action-Sigma-J}), it is described in the Einstein frame as
\begin{align}
I_{\rm E}=&\int_{\cal M_+} \D^{n}x_+ \sqrt{-{\bar g}^+}\biggl(\frac{1}{2\kappa_n}{\bar R}^+-\frac{1}{2}(\nabla\psi^+)^2 -{\bar V}(\psi^+)+{\bar {\cal L}}_{\cal M_+}^{(m)} \biggl) \nonumber \\
&+\int_{\cal M_-} \D^{n}x_- \sqrt{-{\bar g}^-}\biggl(\frac{1}{2\kappa_n}{\bar R}^--\frac{1}{2}(\nabla\psi^-)^2 -{\bar V}(\psi^-)+{\bar {\cal L}}_{\cal M_-}^{(m)} \biggl)  \nonumber \\
&+\frac{\epsilon_+}{\kappa_n} \int_{\partial{\cal M}_+-\Sigma_+} \D^{n-1}z_+ \sqrt{|{\bar \zeta}^+|}{\bar K}^+ +\frac{\epsilon_-}{\kappa_n} \int_{\partial{\cal M}_--\Sigma_-} \D^{n-1}z_- \sqrt{|{\bar \zeta}^-|}{\bar K}^- \nonumber \\
&+\frac{\varepsilon}{\kappa_n} \int_{\Sigma_+} \D^{n-1}y \sqrt{|{\bar h}|}{\bar K}^++\frac{\varepsilon}{\kappa_n} \int_{\Sigma_-} \D^{n-1}y \sqrt{|{\bar h}|}{\bar K}^- +\int_{\Sigma} \D^{n-1}y \sqrt{|{\bar h}|}{\bar {\cal L}}_{\Sigma}^{(m)},\label{action-Sigma-E}
\end{align}
where $\epsilon_+$, $\epsilon_-$, and $\varepsilon$ independently take their values $\pm 1$ and $\psi^\pm|_{\Sigma}=\psi$. 
We assume that the bulk energy-momentum tensor (\ref{bulkmatter-T-E}) do not contain the $\delta$-function part such that
\begin{align}
{\bar T}_{\mu\nu}=\Theta(l){\bar T}^+_{\mu\nu}+\Theta(-l){\bar T}^-_{\mu\nu}.\label{T-decomp-E}
\end{align}

Under the assumptions that ${\bar {\cal L}}_{\cal M_\pm}^{(m)}$ and ${\bar {\cal L}}_{\Sigma}^{(m)}$ do not depend on $\psi^\pm$ and $\psi$, respectively, variation of the above action, with the boundary condition $\delta {\bar g}^\pm_{\mu\nu}|_{\partial{\cal M}_{\pm}-\Sigma_{\pm}}=0=\delta \psi^\pm|_{\partial{\cal M}_{\pm}-\Sigma_{\pm}}$, provides the Einstein equations (\ref{EFE-E}) and the equation of motion for the scalar field (\ref{EOM-E}) in the bulk spacetimes ${\cal M_+}$ and ${\cal M_-}$, as well as, the following junction conditions on $\Sigma$:
\begin{align}
-\varepsilon\left([{\bar K}_{\mu\nu}]-{\bar h}_{\mu\nu}[{\bar 
K}]\right)&=\kappa_n{\bar t}_{\mu \nu},\label{E-frame-j1}\\
{\bar M}&=0,\label{E-frame-j2}
\end{align}
where ${\bar M}:=\varepsilon {\bar n}^\mu[\partial_\mu \psi]$.
${\bar t}_{\mu\nu}$ in Eq.~(\ref{E-frame-j1}) is given by ${\bar t}_{\mu\nu}={\bar t}_{ab}e^a_\mu e^b_\mu$, where
\begin{align}
{\bar t}_{ab}:=&-2\frac{\partial {\bar {\cal L}}_{\Sigma}^{(m)}}{\partial {\bar h}^{ab}}+{\bar h}_{ab}{\bar {\cal L}}_{\Sigma}^{(m)}.
\end{align}

From the junction conditions (\ref{E-frame-j1}) and (\ref{E-frame-j2}), 
one can easily show the following proposition.
\begin{Prop}
\label{Prop:C1matching-E}
({\it E-regularity and E-vacuum at non-null $\Sigma$.})
Suppose in the Einstein frame that \\
(i) $[{\bar g}_{\mu\nu}]=[\psi]=0$ holds at a non-null hypersurface $\Sigma$, and\\
(ii) the second junction conditions at $\Sigma$ are given by Eqs.~(\ref{E-frame-j1}) and (\ref{E-frame-j2}).\\
Then, the $C^1$ regularity at $\Sigma$ is equivalent to ${\bar t}_{ab}\equiv 0$.
\end{Prop}
{\it Proof}.
By Lemma~\ref{lm:regularity-non-null}, the $C^1$ regularity at $\Sigma$ is equivalent to $[{\bar K}_{ab}]=0$.
Then, the proposition follows from Eq.~(\ref{E-frame-j1}) and its trace.
\qed

\section{Junction conditions for null hypersurfaces}
\label{sec;null}


\subsection{Setup}
\begin{figure}[htbp]
\begin{center}
\includegraphics[width=0.8\linewidth]{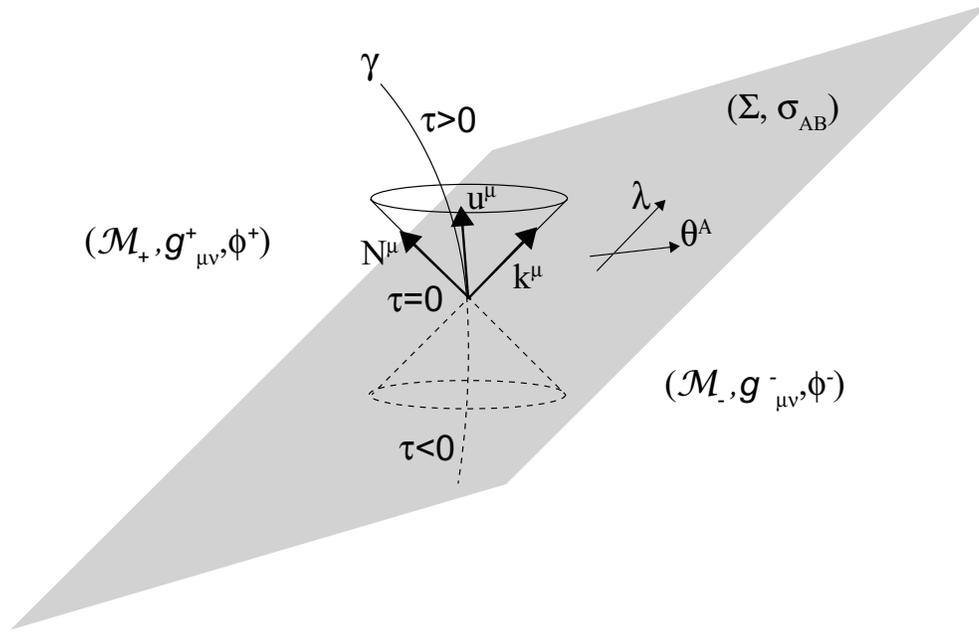}
\caption{\label{Fig-Null-hypersurface} A null hypersurface $\Sigma$ divides the spacetime into two regions ${\cal M}_+$ and ${\cal M}_-$.}
\end{center}
\end{figure}

In the case where $\Sigma$ is null, our convention is such that ${\cal M}_-$ is in the past of $\Sigma$, and ${\cal M}_+$ is in the future. The coordinates $y^a=(\lambda, \theta^A)$  on $\Sigma$  are assumed to be the same 
on both sides of $\Sigma$. We take $\lambda$ to be an arbitrary parameter on the null generators of $\Sigma$ and the other $n-2$ coordinates $\theta^A$ are introduced to label the generators. (See Fig.~\ref{Fig-Null-hypersurface}.)
As for non-null $\Sigma$ in the previous section, we assume that continuous canonical coordinates $x^\mu$, distinct from $x_\pm^\mu$, can be introduced in an open region containing both sides of $\Sigma$.
Hereafter we will describe geometrical and physical quantities in terms of the canonical coordinates $x^\mu$.

The tangent vectors $e^\mu_{a}:=\partial x^\mu/\partial y^a$ on each side of $\Sigma$ are
 naturally separated into a null vector $k^\mu$ that is tangent to the generators, and $n-2$ spacelike vectors $e^\mu_{A}$ that point in the directions transverse to the generators.
$k^\mu$ and $e^\mu_{A}$ are written as
\begin{eqnarray}
k^\mu:=e^\mu_{\lambda}=\biggl(\frac{\partial x^\mu}{\partial \lambda}\biggl)_{\theta^A},~~~~e^\mu_{A}=\biggl(\frac{\partial x^\mu}{\partial \theta^A}\biggl)_{\lambda},
\end{eqnarray}
which satisfy
\begin{eqnarray}
k^\mu k_{\mu}=0=k_{\mu}e^\mu_{A},
\end{eqnarray}
In the canonical coordinates $x^\mu$, both $k^\mu$ and $e^\mu_{A}$ are continuous across $\Sigma$ and hence we have $[k^\mu]=[e^\mu_A]=0$.
The remaining inner products
\begin{eqnarray}
\sigma_{\pm AB}(\lambda,\theta^C):=g^\pm_{\mu\nu}e^\mu_{A}e^\mu_{B}
\end{eqnarray}
are non-vanishing, and we assume that they are also continuous across $\Sigma$:
\begin{eqnarray}
[\sigma_{AB}]:=\sigma_{+ AB}-\sigma_{- AB}=0. \label{sigma}
\end{eqnarray}
The $(n-2)$-tensor $\sigma_{AB}:=\sigma_{+ AB} (\equiv \sigma_{- AB})$ is the induced metric on $\Sigma$:
\begin{eqnarray}
\D s_\Sigma^2=\sigma_{AB}\D \theta^A \D \theta^B.
\end{eqnarray}
The condition (\ref{sigma}) ensures that the intrinsic geometry on $\Sigma$ is well-defined.
The basis in completed by adding an auxiliary null vector $N^\mu$ which satisfies
\begin{eqnarray}
N^\mu N_{\mu}=0,~~~~N^\mu k_{\mu}=-1,~~~~N_{\mu}e^\mu_{A}=0, \label{complete}
\end{eqnarray}
and hence $ N^\mu$ is continuous across $\Sigma$.
The completeness relations of the basis are given as
\begin{eqnarray}
g^{\mu\nu}=-k^\mu N^\nu-N^\mu k^\nu+\sigma^{AB} e^\mu_{A}e^\nu_{B}, \label{comp}
\end{eqnarray}
where the inverse metric $\sigma^{AB}$ on $\Sigma$ is the inverse of $\sigma_{AB}$.

We introduce a congruence of timelike geodesics $\gamma$ that arbitrarily intersect $\Sigma$, of which tangent vector is $u^\mu$. 
Geodesics are parametrized by proper time $\tau$, which is adjusted so that $\tau=0$ at $\Sigma$, $\tau<0$ in ${\cal M}_-$, and $\tau>0$ in ${\cal M}_+$.
Then, the metric $g_{\mu\nu}$ and the scalar field $\phi$ are expressed as distribution-valued tensors in the canonical coordinates $x^\mu$ as
\begin{align}
g_{\mu\nu}=&\Theta(\tau)g_{\mu\nu}^++\Theta(-\tau)g_{\mu\nu}^-, \label{g-n2} \\
\phi=&\Theta(\tau)\phi^++\Theta(-\tau)\phi^-.\label{scalar-n2} 
\end{align}
A displacement along a member of the congruence is described by
\begin{eqnarray}
\D x^\mu=u^\mu\D \tau, \label{dx}
\end{eqnarray}
which is continuous across $\Sigma$, namely $[u^\mu]=0$.
The hypersurface $\Sigma$ is described by $\tau(x^\mu)=0$ and its normal vector $k_{\mu}$ is proportional to the gradient of $\tau(x^\mu)$ evaluated at $\Sigma$.
Hence, the expression of $k_{\mu}$ compatible with Eq.~(\ref{dx}) is
\begin{align}
k_{\mu}=-(-k_{\nu}u^\nu)\frac{\partial \tau}{\partial x^\mu}.\label{k}
\end{align}

\subsection{Continuity of $g_{\mu\nu}$ and $\phi$: First junction conditions}

Differentiating Eqs.~(\ref{g-n2}) and (\ref{scalar-n2}) and using Eq.~(\ref{k}), we obtain
\begin{align}
\partial_\rho g_{\mu\nu}=&\Theta(\tau)\partial_\rho g_{\mu\nu}^++\Theta(-\tau)\partial_\rho g_{\mu\nu}^--(-k_{\eta}u^\eta)^{-1}[g_{\mu\nu}] k_{\rho}\delta(\tau), \label{dg-n2}\\
\partial_\mu \phi=&\Theta(\tau)\partial_\mu \phi^++\Theta(-\tau)\partial_\mu \phi^--(-k_{\eta}u^\eta)^{-1}[\phi] k_\rho\delta(\tau).\label{dscalar-n2} 
\end{align}
As in the case where $\Sigma$ is non-null, to removed the $\delta$ pieces appearing in the right-hand sides, which generate terms proportional to $\Theta(\tau)\delta(\tau)$, we impose $[g_{\mu\nu}]=[\phi]=0$, namely continuity of the metric and $\phi$ across $\Sigma$, and then we have
\begin{align}
\partial_\rho g_{\mu\nu}=&\Theta(\tau)\partial_\rho g_{\mu\nu}^++\Theta(-\tau)\partial_\rho g_{\mu\nu}^-, \label{dg-n2-2}\\
\partial_\mu \phi=&\Theta(\tau)\partial_\mu \phi^++\Theta(-\tau)\partial_\mu \phi^-.\label{dscalar-n2-2} 
\end{align}

Now we characterize the discontinuous behaviors of $\partial_\rho g_{\mu\nu}$ and $\partial_\mu \phi$.
The continuity conditions on the fields guarantee that the tangential derivatives of the metric and scalar field are also continuous, namely
\begin{align}
&[\partial_\rho g_{\mu\nu}]k^\rho=0=[\partial_\rho g_{\mu\nu}]e^\rho_{A},\\
&[\partial_\mu \phi]k^\mu=0=[\partial_\mu \phi]e^\mu_{A}.
\end{align}
The only possible discontinuity is therefore in $N^\rho \partial_\rho g_{\mu\nu}$ and $N^\mu \partial_\mu \phi$, namely the transverse derivatives.
In view of Eq.~(\ref{complete}), there exist a tensor field $\gamma_{\mu\nu}$ and a scalar field $W$ such that
\begin{align}
[\partial_\rho g_{\mu\nu}]=-\gamma_{\mu\nu}k_\rho,\qquad [\partial_\mu \phi]=W k_\mu.\label{ds2-n2}
\end{align}
Namely, $\gamma_{\mu\nu}$ and $W$ are defined by 
\begin{equation} \label{disc2-n2}
\gamma_{\mu\nu}:=N^\rho[\partial_\rho g_{\mu\nu}], \qquad W:=-N^\mu[\partial_\mu \phi],
\end{equation}
respectively.

\subsection{Discontinuity of geometric and physical quantities}

We have seen $[k^\mu]=[e^\mu_A]=[N^\mu]=[u^\mu]=0$ in the canonical coordinates $x^\mu$.
Differentiation of the metric proceeds as in the non-null case, except that we now write $\tau$ instead of $l$, and we use Eq.~(\ref{k}) to relate the gradient of $\tau$ to the null vector $k^\mu$. Thus, we obtain a 
Riemann tensor that contains a singular part given by 
\begin{align}
	R^\rho_{~\sigma\mu\nu}&=\Theta(\tau){R^+}^{\rho}_{~\sigma\mu\nu}+\Theta(-\tau){R^-}^{\rho}_{~\sigma\mu\nu}+\delta(\tau){\tilde R}^\rho_{~\sigma\mu\nu},\label{riemann2} \\
	{\tilde R}^\rho_{~\sigma\mu\nu}&:=-(-k_{\eta}u^\eta)^{-1}([\Gamma^{\rho}_{\sigma\nu}]k_\mu-[\Gamma^{\rho}_{\sigma\mu}]k_\nu). \label{riemann2-deltapart}
\end{align}
where $[\Gamma^{\rho}_{\sigma\nu}]$ is the jump in the Christoffel symbols across $\Sigma$.

Equation~(\ref{ds2-n2}) implies
\begin{align} \label{Gamma-gamma-null}
[\Gamma^\rho_{\sigma\mu}]=-\frac12 (\gamma^\rho_{~~\sigma}k_{\mu}+\gamma^\rho_{~~\mu}k_{\sigma}-\gamma_{\sigma\mu}k^{\rho}),
\end{align}
so that the $\delta$-function part of the Riemann tensor can be written as
\begin{align}
{\tilde R}^\rho_{~\sigma\mu\nu}&=\frac12 (-k_{\eta}u^\eta)^{-1}(\gamma^\rho_{~~\nu}k_{\sigma}k_{\mu}-\gamma^\rho_{~~\mu}k_{\sigma}k_{\nu}-\gamma_{\sigma\nu}k^{\rho}k_{\mu}+\gamma_{\sigma\mu}k^{\rho}k_{\nu}).\label{dR}
\end{align}
We see that $k^\mu$ and $\gamma_{\mu\nu}$ give a complete characterization of the singular part of the Riemann tensor, and the $\delta$-function terms of the Ricci tensor and the Ricci scalar are easily determined as
\begin{align}
{\tilde R}_{\sigma\nu}&=\frac12 (-k_{\eta}u^\eta)^{-1}(\gamma_{\mu\nu}k^{\mu}k_{\sigma}+\gamma_{\mu\sigma}k^{\mu}k_{\nu}-\gamma k_{\sigma}k_{\nu}),\label{A2-n} \\
{\tilde R}&=(-k_{\eta}u^\eta)^{-1}\gamma_{\mu\nu}k^{\mu}k^{\nu},\label{A3-n}
\end{align}
respectively, where $\gamma:=\gamma^\mu_{~~\mu}$.
Finally, the singular part of the Einstein tensor is given by
\begin{align}
{\tilde G}^{\mu\nu}&=\frac12 (-k_{\eta}u^\eta)^{-1}(\gamma_{~~\rho}^{\nu}k^{\rho}k^{\mu}+\gamma_{~~\rho}^{\mu}k^{\rho}k^{\nu}- \gamma k^{\mu}k^{\nu}-g^{\mu\nu}\gamma_{\rho\sigma}k^{\rho}k^{\sigma}). \label{bar-G-n}
\end{align}
The factor $-k_{\eta}u^\eta$ depends on the choice of observers correponding to $u^\mu$ who makes measurements on the shell.

On the other hand, differentiating Eq.~(\ref{dscalar-n2-2}), we obtain
\begin{align}
\partial_\mu\partial_\nu \phi=&\Theta(\tau)\partial_\mu\partial_\nu \phi^++\Theta(-\tau)\partial_\mu\partial_\nu \phi^--(-k_\eta u^\eta)^{-1}W k_\mu k_\nu \delta(\tau), \label{ddscalar2-n} 
\end{align}
where we used Eqs.~(\ref{k}) and (\ref{ds2-n2}).
From the above expression, we get
\begin{align}
\nabla_\mu\nabla_\nu \phi=&\Theta(\tau)\nabla_\mu\nabla_\nu  \phi^++\Theta(-\tau)\nabla_\mu\nabla_\nu  \phi^--(-k_\eta u^\eta)^{-1}W k_\mu k_\nu \delta(\tau), \label{ddscalar2-c-n} \\
\Box \phi=&\Theta(\tau)\Box \phi^++\Theta(-\tau)\Box \phi^-. \label{ddscalar2-c-2-n}
\end{align}
Finally, using the following expression
\begin{align}
\nabla_\mu\nabla_\nu f(\phi)=&f'(\phi)\nabla_\mu\nabla_\nu \phi+f''(\phi)(\nabla_\mu \phi)(\nabla_\nu \phi),
\end{align}
we find
\begin{align}
\nabla_\mu\nabla_\nu f(\phi)=&\Theta(\tau)\nabla_\mu\nabla_\nu f(\phi^+)+\Theta(-\tau)\nabla_\mu\nabla_\nu f(\phi^-)-(-k_\eta u^\eta)^{-1}f'(\phi) W k_\mu k_\nu \delta(\tau), \label{ddf-n} \\
\Box f(\phi)=&\Theta(\tau)\Box f(\phi^+)+\Theta(-\tau)\Box f(\phi^-). \label{Box-f-n}
\end{align}

For later use, we introduce the projections
\begin{align}
\gamma_A:=\gamma_{\mu\nu}e^\mu_A k^\nu,~~~~~\gamma_{AB}:=\gamma_{\mu\nu}e^\mu_A e^\nu_B.
\end{align}
By the completeness relation (\ref{comp}), the vector $\gamma_{\mu\nu}k^\nu$ admits the following decomposition:
\begin{align} \label{gammak}
\gamma_{\mu\nu}k^\nu=\frac12 (\gamma-\sigma^{AB}\gamma_{AB})k_\mu+(\sigma_{AB}\gamma^B)e_\mu^A-(\gamma_{\rho\sigma}k^\rho k^\sigma)N_\mu.
\end{align}
For a consistency check, from the above expression, we obtain $\gamma_{\mu\nu}k^\nu N_\mu=(1/2)\gamma_{\mu\nu}(k^\mu N^\nu +k^\nu N^\mu)=\gamma_{\mu\nu}k^\mu N^\nu$, where the last equality holds because of the symmetric nature of $\gamma_{\mu\nu}$.

Since $k^\mu$ is not normal but tangent to $\Sigma$, we introduce a transverse curvature $C_{ab}$ that properly represents the transverse derivative of the metric:
\begin{align}
C_{ab}&:=\frac12 ({\cal L}_N g_{\mu\nu})e^\mu_a e^\nu_b=(\nabla_\mu N_{\nu})e^\mu_{a}e^\nu_b, 
\end{align}
where we have used that $N_\mu e^\mu_a=0$ and an identity $(\nabla_\nu e^\mu_{a})e^\nu_b\equiv (\nabla_\nu e^\mu_{b})e^\nu_{a}$.
In the canonical coordinates $x^\mu$, the jump of the transverse curvature at $\Sigma$ is given by
\begin{align}
[C_{ab}]&=[\nabla_\mu N_{\nu}]e^\mu_{a}e^\nu_b=\frac12 \gamma_{\mu\nu}e^\mu_{a}e^\nu_b. \label{Cab-jump}
\end{align}
We therefore have
\begin{equation}
[C_{\lambda\lambda}]=\frac12 \gamma_{\mu\nu}k^\mu k^\nu,~~~[C_{A\lambda}]=\frac12 \gamma_{A},~~~[C_{AB}]=\frac12 \gamma_{AB}.\label{relations-n}
\end{equation}

As in Lemma~\ref{lm:regularity-non-null} for non-null $\Sigma$, the following lemma provides several different expressions of a $C^1$ regular matching condition, $[g_{\alpha \beta}]=[\partial_\mu g_{\alpha \beta}]=0$, in the case where $\Sigma$ is null.
\begin{lm}
\label{lm:regularity-null}
If $[g_{\alpha \beta}]=0$ holds, the following five conditions are equivalent: (i) ${\tilde R}^\rho_{~\sigma\mu\nu}=0$, (ii) $[C_{ab}]=0$, (iii) $\gamma_{\mu\nu}=0$, (iv) $[\partial_\mu g_{\alpha \beta}]=0$, and (v) $[\Gamma^\rho_{\sigma\mu}]=0$.
\end{lm}
{\it Proof:}
The conditions (ii) and (iii) are equivalent by Eq.~(\ref{Cab-jump}).
The conditions (iii) and (iv) are equivalent by Eqs.~(\ref{ds2-n2}) and (\ref{disc2-n2}).
Now we show that the conditions (i) and (ii) are equivalent.
If $[C_{ab}]=0$ holds, we have $\gamma_{\mu\nu}=0$ and hence ${\tilde R}^\rho_{~\sigma\mu\nu}=0$ is satisfied by Eq.~(\ref{dR}).
On the other hand, if ${\tilde R}^\rho_{~\sigma\mu\nu}=0$ holds, Eq.~(\ref{dR}) gives
\begin{align}
\gamma_{\rho\nu}k_{\sigma}k_{\mu}-\gamma_{\rho\mu}k_{\sigma}k_{\nu}-\gamma_{\sigma\nu}k_{\rho}k_{\mu}+\gamma_{\sigma\mu}k_{\rho}k_{\nu}=0.
\end{align}
Acting $k^\rho k^\nu$, $k^\rho e^\nu_A$, and $e^\rho_A e^\nu_B$ on the above equation, we respectively obtain $\gamma_{\rho\nu}k^{\rho}k^{\nu}=0$, $\gamma_{\rho\nu}k^{\rho} e^\nu_A=0$, and $\gamma_{\rho\nu}e^\rho_Ae^\nu_B=0$, and hence $[C_{ab}]=0$ is concluded. 
Since we have shown that the conditions (i)--(iv) are equivalent, we complete the proof by showing that the conditions (iii) and (v) are equivalent.
The condition (iii) implies the condition (v) by Eq.~(\ref{Gamma-gamma-null}).
The condition (v) implies the condition (i) by Eq.~(\ref{riemann2-deltapart}), which is equivalent to the condition (iii).
\qed

\bigskip

\subsection{Second junction conditions}
\subsubsection{Equation of motion for a scalar field}
Here we derive the junction condition from the equation of motion (\ref{EOM-J}), namely $\Pi=0$, where $\Pi$ is defined by Eq.~(\ref{EOM}).
Using Eqs.~(\ref{A3-n}) and (\ref{ddscalar2-c-2-n}), we write down $\Pi$ as
\begin{align} \label{equationKGnull}
\Pi=\Theta(\tau)\Pi^++\Theta(-\tau)\Pi^-+\delta (\tau){\tilde\Pi},
\end{align}
where the $\delta$-function part ${\tilde \Pi}$ is given by 
\begin{align} 
{\tilde \Pi}:=&(-k_{\eta}u^\eta)^{-1}f'(\phi)\gamma_{\mu\nu}k^{\mu}k^{\nu}. \label{barPi-def-n}
\end{align}
Thus, the equation of motion $\Pi=0$ on $\Sigma$ gives ${\tilde\Pi}=0$, namely
\begin{equation} \label{Mreg-n}
f'(\phi)\gamma_{\mu\nu}k^{\mu}k^{\nu}=0.
\end{equation}
We shall refer to this condition as the junction condition from the equation of motion for a scalar field.
For a minimally coupled scalar field, namely for $f(\phi)=1/(2\kappa_n)$, this condition is trivially satisfied.

\subsubsection{Einstein equations}
Next let us derive the junction conditions from the Einstein equations (\ref{EFE-J}), namely $E_{\mu\nu}=T_{\mu\nu}$, where $E_{\mu\nu}$ is defined by Eq.~(\ref{EFE}).
Using Eqs.~(\ref{bar-G-n}), (\ref{ddf-n}), and (\ref{Box-f-n}), we write down $E_{\mu\nu}$ as
\begin{align}
E_{\mu\nu}=&\Theta(\tau){E}^+_{\mu\nu}+\Theta(-\tau){E}^-_{\mu\nu}+\delta(\tau){\tilde E}_{\mu\nu},\label{E-decomp-n}
\end{align}
where, the $\delta$-function part ${\tilde E}_{\mu\nu}$ is given by 
\begin{align}
{\tilde E}_{\mu\nu}:=&(-k_{\eta}u^\eta)^{-1} \nonumber \\
&\times\biggl\{f(\phi)(\gamma_{\nu\rho}k^{\rho}k_{\mu}+\gamma_{\mu\rho}k^{\rho}k_{\nu}-g_{\mu\nu}\gamma_{\rho\sigma}k^{\rho}k^{\sigma})+\left(2f'(\phi) W-f(\phi)\gamma\right) k_\mu k_\nu\biggl\}.
\end{align}
We assume that the bulk matter fields do not contain the $\delta$-function part such that
\begin{align}
T_{\mu\nu}=\Theta(\tau){T}^+_{\mu\nu}+\Theta(-\tau){T}^-_{\mu\nu},\label{T-decomp-null}
\end{align}
which means that the bulk matter fields do not contribute to the energy-momentum tensor on $\Sigma$.
By Eqs.~(\ref{E-decomp-n}) and (\ref{T-decomp-null}), the Einstein equations $E_{\mu\nu}=T_{\mu\nu}$ on $\Sigma$ give ${\tilde E}_{\mu\nu}=0$, which we shall refer as the junction conditions from the Einstein equations, which are the conditions for vacuum $\Sigma$.

For embedding configurations of $\Sigma$ with ${\tilde E}_{\mu\nu}\ne 0$, the Einstein equations require an additional matter field on $\Sigma$ for consistency, so that $\Sigma$ is no more vacuum. 
The Einstein equations on $\Sigma$ then become
\begin{align}
{\tilde E}_{\mu\nu}=t_{\mu \nu}, \label{Tthin-null}
\end{align}
where $t_{\mu \nu}$ is the thin-shell energy-momentum tensor on $\Sigma$, which is written as
\begin{align}
t_{\mu \nu}=&(-k_{\eta}u^\eta)^{-1} \nonumber \\
&\times\biggl\{f(\phi)(\gamma_{\nu\rho}k^{\rho}k_{\mu}+\gamma_{\mu\rho}k^{\rho}k_{\nu}-g_{\mu\nu}\gamma_{\rho\sigma}k^{\rho}k^{\sigma})+\left(2f'(\phi) W-f(\phi)\gamma\right) k_\mu k_\nu\biggl\}.
\end{align}
The expression of $t_{\mu\nu}$ can be simplified if we decompose it in the basis $\{k^\mu,e^\mu_A,N^\mu\}$.
Using Eq.~\eqref{gammak} and involving once more the completeness relation (\ref{comp}),  $t_{\mu\nu}$ is written as
\begin{align}
t_{\mu \nu}= (-k_{\eta}u^\eta)^{-1}\left\{\mu k_\mu k_\nu +j_A(k_\mu e_{\nu}^A+e_{\mu}^A k_\nu)+p\sigma_{AB}e_\mu^A e_\nu^B\right\} \label{junc-null-t}
\end{align}
with
\begin{align}
\mu&:=2f'(\phi) W-f(\phi)\sigma^{AB}\gamma_{AB},\label{components1-n}\\
j_A&:=f(\phi)\sigma_{AB}\gamma^B,\label{components2-n}\\
p&:=-f(\phi)\gamma_{\mu\nu}k^\mu k^\nu,\label{components3-n}
\end{align}
where $\mu$, $j^A$, and $p$ are respectively interpreted as the shell's surface density, a surface current, and an isotropic surface pressure in the Einstein frame~\cite{Poisson:2002nv}.
By Eq.~(\ref{relations-n}), the surface quantities  \eqref{components1-n}--\eqref{components3-n} can be expressed in terms of the transverse curvature such that
\begin{align}
\mu&=2f'(\phi) W-2f(\phi)\sigma^{AB}[C_{AB}],\label{components1-n2}\\
j^A&=2f(\phi)\sigma^{AB}[C_{\lambda B}],\label{components2-n2}\\
p&=-2f(\phi)[C_{\lambda\lambda}]. \label{components3-n2}
\end{align}

In summary, we have obtained the junction conditions (\ref{Mreg-n}) and (\ref{junc-null-t}) in the Jordan frame at a null hypersurface $\Sigma$  as
\begin{align}
&t_{\mu \nu}= (-k_{\eta}u^\eta)^{-1}\left\{\mu k_\mu k_\nu +j_A(k_\mu 
e_{\nu}^A+e_{\mu}^A k_\nu)+p\sigma_{AB}e_\mu^A e_\nu^B\right\}, 
\label{null-t-J1}\\
&f'(\phi)[C_{\lambda\lambda}]=0,\label{null-t-J2}
\end{align}
where $\mu$, $j^A$, and $p$ are defined by Eqs.~\eqref{components1-n}--\eqref{components3-n} (or equivalently Eqs.~\eqref{components1-n2}--\eqref{components3-n2}) and $\gamma_{\mu\nu}$ and $W$ are defined by Eq.~(\ref{disc2-n2}).

\subsection{Conditions for $C^1$ matching and vacuum $\Sigma$}
\subsubsection{Jordan frame}
Now let us study the conditions for a $C^1$ matching at $\Sigma$ and also for vacuum $\Sigma$.
\begin{Prop}
\label{Prop:C1matching-null}
({\it J-regularity at null $\Sigma$.})
Let $\phi_\Sigma$ be the value of $\phi$ at a null hypersurface $\Sigma$.
Suppose in the Jordan frame that\\
(i) $f(\phi)$ is a $C^1$ function, \\
(ii) $[\sigma_{AB}]=[\phi]=0$ holds at $\Sigma$, and\\
(iii) the second junction conditions at $\Sigma$ are given by Eqs.~(\ref{null-t-J1}) and (\ref{null-t-J2}). \\
Then, the $C^1$ regularity at $\Sigma$ implies $j_A=p=0$ and $\mu=2f'(\phi_\Sigma) W$.
If $f(\phi_\Sigma)\ne 0$ holds, $j_A=p=0$ and $\mu=2f'(\phi_\Sigma) W$ at $\Sigma$ imply $[C_{\lambda\lambda}]=[C_{\lambda A}]=\sigma^{AB}[C_{AB}]=0$.
\end{Prop}
{\it Proof}.
By Lemma~\ref{lm:regularity-null}, the $C^1$ regularity at $\Sigma$ is equivalent to $[C_{ab}]=0$.
Then, the proposition follows from Eqs.~(\ref{components1-n2})--(\ref{components3-n2}).
\qed

The above proposition suggests a possibility of a $C^1$ matching at $\Sigma$ with non-vanishing $\mu$ if $f'(\phi_\Sigma) W\ne 0$ holds.
This {\it non-vacuum $C^1$ matching} is characteristic in the Jordan frame and clearly shows that $[C_{ab}]=0$ and $t_{\mu\nu}=0$ are not equivalent in this frame.
We also note that $\sigma^{AB}[C_{AB}]=0$ is a weaker condition than $[C_{AB}]=0$.

Now let us obtain conditions for vacuum $\Sigma$ ($t_{\mu\nu}\equiv 0$) in the case where $\Sigma$ is null.
\begin{Prop}
\label{Prop:C1-null-J}
({\it J-vacuum at null $\Sigma$.})
Let $\phi_\Sigma$ be the value of $\phi$ at a null hypersurface $\Sigma$.
Then, under the assumptions (i)--(iii) in Proposition~\ref{Prop:C1matching-null}, $t_{\mu\nu}\equiv 0$ is realized at 
$\Sigma$ only in one of the following three cases:\\
(I) $f(\phi_\Sigma)=0$ and $f'(\phi_\Sigma)=0$,  \\
(II) $f(\phi_\Sigma)=0$ and $[C_{\lambda\lambda}]=W=0$, or \\
(III) $[C_{\lambda\lambda}]=0$, $f'(\phi_\Sigma) 
W=f(\phi_\Sigma)\sigma^{AB}[C_{AB}]$, and $[C_{\lambda A}]=0$.
\end{Prop}
{\it Proof}.
With $t_{\mu\nu}=0$, Eqs.~(\ref{null-t-J1}) and (\ref{null-t-J2}) reduce to
\begin{align} 
&f'(\phi) W=f(\phi)\sigma^{AB}[C_{AB}],\label{junc-n2}\\
&f(\phi)\sigma^{AB}[C_{\lambda B}]=0,\label{junc-n3}\\
&f(\phi)[C_{\lambda\lambda}]=0,\label{junc-n4}\\
&f'(\phi)[C_{\lambda\lambda}]=0.\label{junc-n1}
\end{align}
The proposition follows from Eqs.~(\ref{junc-n2})--(\ref{junc-n1}).
\qed

As discussed in \cite{Barrabes:1991ng}, $\gamma_{\mu\nu}k^{\mu}k^{\nu}(\equiv 2[C_{\lambda\lambda}])=0$ is satisfied for an affine parametrization of the generators of $\Sigma$, which is shown as
\begin{align} 
[C_{\lambda\lambda}]=[\nabla_\mu N_{\nu}]e^\mu_{\lambda}e^\nu_{\lambda}=[\nabla_\mu N_{\nu}] k^\mu k^\nu=-N_{\nu} k^\mu [\nabla_\mu k^\nu]=0.
\end{align}
However, this parametrization is not always possible because in general $\lambda$ cannot be an affine parameter on both sides of $\Sigma$. (See section 3.11.5 in the textbook~\cite{Poissonbook}.)

\subsubsection{Einstein frame}
In the Einstein frame, we assume that the energy-momentum tensor of the bulk matter fields do not contain the $\delta$-function part such that
\begin{align}
{\bar T}_{\mu\nu}=\Theta(\tau){\bar T}^+_{\mu\nu}+\Theta(-\tau){\bar T}^-_{\mu\nu},\label{T-decomp-null-E}
\end{align}
which means that the bulk matter fields do not contribute to the energy-momentum tensor ${\bar t}_{\mu\nu}$ on $\Sigma$.
Since the junction condition (\ref{Mreg-n}) is trivially satisfied, the junction conditions in the Einstein frame are
\begin{align}
&{\bar t}_{\mu \nu}= (-{\bar k}_{\eta}{\bar u}^\eta)^{-1}\biggl\{{\bar 
\mu} {\bar k}_\mu {\bar k}_\nu +{\bar j}_A({\bar k}_\mu 
e_{\nu}^A+e_{\mu}^A {\bar k}_\nu)+{\bar p}{\bar \sigma}_{AB}e_\mu^A 
e_\nu^B\biggl\}, \label{null-t-E}
\end{align}
where
\begin{align}
{\bar \mu}&:=-\kappa_n^{-1}{\bar \sigma}^{AB}[{\bar C}_{AB}],\label{components1-n2-Eframe}\\
{\bar j}^A&:=\kappa_n^{-1}{\bar \sigma}^{AB}[{\bar C}_{\lambda B}],\label{components2-n2-Eframe}\\
{\bar p}&:=-\kappa_n^{-1}[{\bar C}_{\lambda\lambda}]. \label{components3-n2-Eframe}
\end{align}
Therefore, ${\bar t}_{\mu\nu}\equiv 0$ at $\Sigma$ is equivalent to 
\begin{align}
{\bar \sigma}^{AB}[{\bar C}_{AB}]=[{\bar C}_{\lambda B}]=[{\bar C}_{\lambda\lambda}]=0. \label{junc-n4-E}
\end{align}
The following proposition clarifies the relation between vacuum $\Sigma$ (${\bar t}_{\mu\nu}\equiv 0$) and a $C^1$ matching at $\Sigma$ in the Einstein frame.
\begin{Prop}
\label{Prop:C1matching-null-E}
({\it E-regularity and E-vacuum at null $\Sigma$.})
Suppose in the Einstein frame that\\
(i) $[{\bar \sigma}_{AB}]=[\psi]=0$ holds at a null hypersurface $\Sigma$, and \\
(ii) the second junction condition at $\Sigma$ is given by Eq.~(\ref{null-t-E}).\\
Then, the $C^1$ regularity at $\Sigma$ implies ${\bar t}_{\mu\nu}=0$.
${\bar t}_{\mu\nu}=0$ at $\Sigma$ implies $[{\bar C}_{\lambda\lambda}]=[{\bar C}_{\lambda A}]={\bar \sigma}^{AB}[{\bar C}_{AB}]=0$.
\end{Prop}
{\it Proof}.
By Lemma~\ref{lm:regularity-null}, the $C^1$ regularity at $\Sigma$ is equivalent to $[{\bar C}_{ab}]=0$.
Then, the proposition follows from Eqs.~(\ref{components1-n2-Eframe})--(\ref{components3-n2-Eframe}).
\qed

While Proposition~\ref{Prop:C1matching-E} shows that ${\bar t}_{\mu\nu}\equiv 0$ and $[{\bar K}_{\mu\nu}]=0$ are equivalent in the case where $\Sigma$ is non-null, Proposition~\ref{Prop:C1matching-null-E} shows that ${\bar t}_{\mu\nu}=0$ is just a necessary condition for $[{\bar C}_{ab}]=0$ in the case where $\Sigma$ is null because ${\bar \sigma}^{AB}[{\bar C}_{AB}]=0$ is weaker than $[{\bar C}_{AB}]=0$.

\section{Relation between the conditions in Jordan and Einstein frames}
\label{sec:equivalence}

In this section, we study the relation between $C^1$ matchings in Jordan and Einstein frames and also the relation between the conditions for vacuum $\Sigma$. 
As seen in Sec.~\ref{sec:proper}, the matter Lagrangian densities may introduce anomalies which 
violate the correspondence between the Jordan and Einstein frames.
Indeed, as shown in Appendix~\ref{App:JtoE} in the case where $\Sigma$ is non-null, there is a proper mapping between the Jordan frame (\ref{action-Sigma-J}) and the Einstein 
frame (\ref{action-Sigma-E}) only when the non-minimal coupling $f(\phi)$ does not satisfy Eq.~(\ref{exceptional}) and the extra matter fields in the bulk and on $\Sigma$ are conformal invariant, namely $T_{\mu\nu}={\bar T}_{\mu \nu}$ and $t_{ab}={\bar t}_{ab}$, including vacuum cases.
Only in such cases, there exists a proper correspondence between the field equations and junction conditions in two frames.

Even if there is no proper correspondence between them, one can study the relation of the $C^1$ regular matchings in the Jordan and Einstein frames because it is a purely geometrical concept.
Naively thinking, the $C^1$ regular matchings in two frames seem to be equivalent; however, we will see that there are some exceptional cases.
We first show the following lemma for later use.
\begin{lm}
\label{lm:first}
Let $\phi_\Sigma$ be the value of $\phi$ at a null or non-null hypersurface $\Sigma$.
If $f(\phi)$ is a $C^1$ function and not in the exceptional form (\ref{exceptional}), then $[\phi]=0$ and $[\psi]=0$ are equivalent.
If $f(\phi_\Sigma)\ne 0$ holds in addition, then $[g_{\mu\nu}]=0$ and $[{\bar g}_{\mu\nu}]=0$ are equivalent.
\end{lm}
{\it Proof:}
Since $f(\phi)$ is in the $C^1$-class and not in the exceptional form (\ref{exceptional}), Eq.~(\ref{def-psi}) with a fixed sign in the right-hand side shows that $\psi(\phi)$ is a continuous and monotonic function.
Thus, there exists a continuous inverse function $\phi(\psi)$ and hence $[\phi]=0$ and $[\psi]=0$ are equivalent.
Since $f(\phi_\Sigma)$ is assumed to be non-zero and finite at $\Sigma$, the relation 
${\bar g}_{\mu\nu}=(2 \kappa_n f(\phi))^{2/(n-2)}g_{\mu\nu}$ shows that 
$[g_{\mu\nu}]=0$ is equivalent to $[{\bar 
g}_{\mu\nu}]=0$.
\qed

In the case of $f(\phi_\Sigma)=0$, the geometric information in the other frame cannot be obtained so that one has to study the other frame individually.
Actually, $f(\phi_\Sigma)=0$ is a part of the J-vacuum condition (II) for non-null $\Sigma$ in Proposition~\ref{Prop:C1-non-null-J} as well as the J-vacuum conditions (I) and (II) for null $\Sigma$ in Proposition~\ref{Prop:C1-null-J}.
In the following subsections, we will see that $2(n-1){f'(\phi_\Sigma)}^2+(n-2)f(\phi_\Sigma)=0$ is also such an exceptional case.

\subsection{Non-null hypersurfaces}

Using Eqs.~(\ref{n-def-E}) and (\ref{def-psi}), we obtain
\begin{align}
\varepsilon {\bar n}^\mu\partial_\mu \psi&=\pm\varepsilon 
(2 \kappa_nf(\phi))^{-1/(n-2)} 
\sqrt{\frac{2(n-1){f'(\phi)}^2+(n-2)f(\phi)}{2(n-2)\kappa_n 
f(\phi)^2}}{n}^\mu\partial_\mu \phi,\label{proof-nonnull1}
\end{align}
while Eq.~(\ref{rel-KK}) gives
\begin{align}
{\bar 
K}_{\mu\nu}=(2 \kappa_nf(\phi))^{1/(n-2)}\left(K_{\mu\nu}+\frac{f'(\phi)}{(n-2)f(\phi)}n^\sigma(\partial_\sigma\phi)h_{\mu\nu}\right).\label{proof-nonnull2}
\end{align}
From the above equations, we first clarify the relations of the $C^1$ regularity at $\Sigma$ in the Jordan and Einstein frames.
\begin{Prop}
\label{Prop:non-null-regularity-relation}
(Relation of $C^1$-regularities at non-null $\Sigma$.)
Let $\phi_\Sigma$ be the value of $\phi$ at a non-null hypersurface $\Sigma$.
Suppose that $f(\phi)$ is a $C^1$ function, not in the exceptional form (\ref{exceptional}), and satisfies $f(\phi_\Sigma)\ne 0$.
Then, \\
(i) under the assumptions in Proposition~\ref{Prop:C1matching} in the Jordan frame, $[K_{\mu\nu}]=0$ implies $[{\bar K}_{\mu\nu}]=0$, and\\
(ii) under the assumptions in Proposition~\ref{Prop:C1matching-E} in the Einstein frame, $[{\bar K}_{\mu\nu}]=0$ implies $[K_{\mu\nu}]=0$ if $2(n-1){f'(\phi_\Sigma)}^2+(n-2)f(\phi_\Sigma)\ne 0$.
\end{Prop}
{\it Proof}.
By Lemma~\ref{lm:first}, $[\phi]=[g_{\mu\nu}]=0$ and $[\psi]=[{\bar g}_{\mu\nu}]=0$ are equivalent.
Then, Eqs~(\ref{proof-nonnull1}) and (\ref{proof-nonnull2}) give
\begin{align}
{\bar M}=&\pm (2 \kappa_nf(\phi))^{-1/(n-2)} 
\sqrt{\frac{2(n-1){f'(\phi)}^2+(n-2)f(\phi)}{2(n-2)\kappa_n 
f(\phi)^2}}M,\label{rel-MM}\\
[{\bar 
K}_{\mu\nu}]=&(2 \kappa_nf(\phi))^{1/(n-2)}\left([K_{\mu\nu}]+\frac{\varepsilon 
f'(\phi)}{(n-2)f(\phi)}M h_{\mu\nu}\right).\label{rel-K-KM}
\end{align}
$[K_{\mu\nu}]=0$ in the Jordan frame implies $M=t_{\mu\nu}=0$ by the junction conditions (\ref{j-eq-summary1}) and (\ref{j-eq-summary2}), which shows $[{\bar K}_{\mu\nu}]={\bar M}=0$ by Eqs.~(\ref{rel-MM}) and (\ref{rel-K-KM}).
On the other hand, $[{\bar K}_{\mu\nu}]=0$ in the Einstein frame implies ${\bar M}={\bar t}_{\mu\nu}=0$ by the junction conditions (\ref{E-frame-j1}) and (\ref{E-frame-j2}), which shows $[K_{\mu\nu}]=M=0$ by Eqs.~(\ref{rel-MM}) and (\ref{rel-K-KM}).
\qed

The above proposition does not assume a proper correspondence between the Jordan and Einstein frames.
If there is, the statement (i) leads ${\bar t}_{\mu\nu}=0$ by Proposition~\ref{Prop:C1matching-E}, while the statement (ii) leads $t_{\mu\nu}=0$ by Proposition~\ref{Prop:C1-non-null-J}.

Now we clarify the relation of the vacuum $\Sigma$ conditions in the case where there is a proper correspondence between two frames.

\begin{Prop}
\label{Prop:non-null}
(Relation of vacuum non-null $\Sigma$.)
Let $\phi_\Sigma$ be the value of $\phi$ at a non-null hypersurface $\Sigma$.
Suppose that \\
(i) $f(\phi)$ is a $C^1$ function, not in the exceptional form (\ref{exceptional}), and satisfies $f(\phi_\Sigma)\ne 0$,\\
(ii) there is a proper correspondence between the Jordan and Einstein frames, and\\
(iii) the assumptions in Proposition~\ref{Prop:C1matching} in the Jordan frame and in Proposition~\ref{Prop:C1matching-E} in the Einstein frame hold.\\
Then, J-vacuum condition (I) or (III) in Proposition~\ref{Prop:C1-non-null-J} implies E-vacuum.
E-vacuum implies the J-vacuum condition (I) if $2(n-1){f'(\phi_\Sigma)}^2+(n-2)f(\phi_\Sigma)\ne 0$ holds.
\end{Prop}
{\it Proof}.
By Eq.~(\ref{rel-K-KM}), both J-vacuum conditions (I) and (III) in Proposition~\ref{Prop:C1-non-null-J} imply ${\bar M}=[{\bar K}_{\mu\nu}]=0$, which shows ${\bar t}_{\mu\nu}=0$ by Proposition~\ref{Prop:C1matching-E}.
By Proposition~\ref{Prop:C1matching-E}, E-vacuum ${\bar t}_{\mu\nu}=0$ is equivalent to $[{\bar K}_{\mu\nu}]=0$, which shows $t_{\mu\nu}=0$ by Proposition~\ref{Prop:non-null-regularity-relation}.
\qed

\subsection{Null hypersurfaces}

While ${\bar u}^\mu=u^\mu$ holds, ${\bar 
g}_{\mu\nu}=(2 \kappa_nf(\phi))^{2/(n-2)}g_{\mu\nu}$ shows that the relations between the pseudo-orthonormal 
basis and the induced metric in the Jordan and Einstein frames are
\begin{align}
&{\bar N}^\mu=(2 \kappa_nf(\phi))^{-1/(n-2)}N^{\mu},\quad {\bar 
k}^\mu=(2 \kappa_nf(\phi))^{-1/(n-2)}k^{\mu}, \label{basis-rel-JE1}\\
&{\bar 
\sigma}^{AB}=(2 \kappa_nf(\phi))^{-2/(n-2)}\sigma^{AB},\quad {\bar e}^\mu_A=e^\mu_A, 
\label{basis-rel-JE2}
\end{align}
which satisfy the following completeness condition in the Einstein frame:
\begin{eqnarray}
{\bar g}^{\mu\nu}=-{\bar k}^\mu{\bar N}^\nu-{\bar N}^\mu{\bar 
k}^\nu+{\bar \sigma}^{AB} {\bar e}^\mu_{A} {\bar e}^\nu_{B}. 
\end{eqnarray}
Now let us clarify the relations of the $C^1$ regularity at $\Sigma$ in the Jordan and Einstein frames.

\begin{Prop}
\label{Prop:null-regularity-relation}
(Relation of $C^1$-regularities at null $\Sigma$.)
Let $\phi_\Sigma$ be the value of $\phi$ at a null hypersurface $\Sigma$.
Suppose that $f(\phi)$ is a $C^1$ function, not in the exceptional form (\ref{exceptional}), and satisfies $f(\phi_\Sigma)\ne 0$. Then, the following two statements hold:\\
(i) $[C_{ab}]=f'(\phi_\Sigma)W=0$ in the Jordan frame implies $[{\bar C}_{ab}]=0$, and\\
(ii) $[{\bar C}_{ab}]={\bar W}=0$ in the Einstein frame implies $[C_{ab}]=0$ if $2(n-1){f'(\phi_\Sigma)}^2+(n-2)f(\phi_\Sigma)\ne 0$.
\end{Prop}
{\it Proof}.
Since $f(\phi_\Sigma)$ is non-zero and finite, Eqs.~(\ref{basis-rel-JE1}) and (\ref{basis-rel-JE2}) show that $[N^{\mu}]=[k^{\mu}]=[\sigma_{AB}]=[e^\mu_{A}]=0$ are equivalent to $[{\bar N}^{\mu}]=[{\bar k}^{\mu}]=[{\bar \sigma}_{AB}]=[{\bar e}^\mu_{A}]=0$.
Also, by Lemma~\ref{lm:first}, $[\phi]=0$ and $[\psi]=0$ are equivalent.
Then, the following relations
\begin{align}
-{\bar N}^\mu\partial_\mu \psi=&-\sqrt{\frac{2(n-1){f'(\phi)}^2+(n-2)f(\phi)}{2(n-2)\kappa_nf(\phi)^2}}(2 \kappa_nf(\phi))^{-1/(n-2)}{N}^\mu\partial_\mu
\phi, \label{rel-EJ-null1}\\
{\bar N}^\rho \partial_\rho {\bar g}_{\mu\nu}=&(2 \kappa_nf(\phi))^{-1/(n-2)}{N}^\rho\partial_\rho ((2 \kappa_nf(\phi))^{2/(n-2)}{g}_{\mu\nu}) \nonumber \\
=&(2 \kappa_nf(\phi))^{1/(n-2)}\left(\frac{2}{n-2}f(\phi)^{-1}f'(\phi)N^\rho(\partial_\rho 
\phi){g}_{\mu\nu}+N^\rho\partial_\rho{g}_{\mu\nu}\right), 
\label{rel-EJ-null2}
\end{align}
give
\begin{align}
{\bar 
W}=&\sqrt{\frac{2(n-1){f'(\phi)}^2+(n-2)f(\phi)}{2(n-2)\kappa_nf(\phi)^2}}(2 \kappa_nf(\phi))^{-1/(n-2)}W
\label{rel-EJ-null1-2}\\
[{\bar 
C}_{ab}]=&(2 \kappa_nf(\phi))^{1/(n-2)}\left(-\frac{1}{n-2}\frac{f'(\phi)}{f(\phi)}W{g}_{\mu\nu}e^\mu_{a}e^\nu_b+[{C}_{ab}]\right),\label{rel-EJ-CC}
\end{align}
where we used $[C_{ab}]=[\nabla_\mu 
N_{\nu}]e^\mu_{a}e^\nu_b=(1/2)N^\rho[\partial_\rho 
g_{\mu\nu}]e^\mu_{a}e^\nu_b$.
The proposition follows from Eqs.~(\ref{rel-EJ-null1-2}) and (\ref{rel-EJ-CC}).
\qed

The above proposition is purely geometrical and does not assume the second junction conditions.
In fact, even with the second junction conditions, the $C^1$ matching conditions at null $\Sigma$ in the Jordan and Einstein frames are not equivalent.

Next we clarify the relations of vacuum $\Sigma$ conditions in the case where there is a proper correspondence between two frames.
\begin{Prop}
\label{Prop:null}
(Relation of vacuum null $\Sigma$.)
Let $\phi_\Sigma$ be the value of $\phi$ at a null hypersurface $\Sigma$.
Suppose that \\
(i) $f(\phi)$ is a $C^1$ function, not in the exceptional form (\ref{exceptional}), and satisfies $f(\phi_\Sigma)\ne 0$,\\
(ii) there is a proper correspondence between the Jordan and Einstein frames, and\\
(iii) the assumptions in Proposition~\ref{Prop:C1matching-null} in the Jordan frame and in Proposition~\ref{Prop:C1matching-null-E} in the Einstein frame hold.\\
Then, the J-vacuum condition (III) in Proposition~\ref{Prop:C1-null-J} is equivalent to E-vacuum.
\end{Prop}
{\it Proof}.
Equation~(\ref{rel-EJ-CC}) gives
\begin{align}
[{\bar C}_{\lambda\lambda}]=&(2 \kappa_nf(\phi))^{1/(n-2)}[{C}_{\lambda\lambda}],\label{C-proof-null1}\\
[{\bar C}_{\lambda B}]=&(2 \kappa_nf(\phi))^{1/(n-2)}[{C}_{\lambda B}],\label{C-proof-null2}\\
[{\bar C}_{AB}]=&(2 \kappa_nf(\phi))^{1/(n-2)}\left(-\frac{1}{n-2}\frac{f'(\phi)}{f(\phi)}W\sigma_{AB}+[{C}_{AB}]\right),\label{C-proof-null3}\\
{\bar \sigma}^{AB}[{\bar C}_{AB}]=&(2 \kappa_nf(\phi))^{-1/(n-2)}\left(-\frac{f'(\phi)}{f(\phi)}W+\sigma^{AB}[{C}_{AB}]\right).\label{C-proof-null4}
\end{align}
E-vacuum (${\bar t}_{\mu\nu}=0$) is equivalent to Eq.~(\ref{junc-n4-E}).
By Eqs.~(\ref{C-proof-null1}), (\ref{C-proof-null2}), and (\ref{C-proof-null4}), Eq.~(\ref{junc-n4-E}) is equivalent to the J-vacuum condition (III) in Proposition~\ref{Prop:C1-null-J}.
\qed

\subsection{Examples of vacuum $C^1$ matching at null hypersurface} \label{examples}
Here we present two examples of the vacuum $C^1$ matching at a null hypersurface.
Since extra matter fields do not exist in the bulk spacetime, there is a proper correspondence in the Jordan and Einstein frames in both cases.

\subsubsection{Roberts-(A)dS solution in the Einstein frame ($n=4$)}

Let us consider the Einstein-$\Lambda$ system with a massless scalar 
field $\phi$ in four dimensions, of which action is given by
\begin{align}
I_{\rm E}=&\int_{\cal M} \D^{4}x \sqrt{-{\bar 
g}}\biggl(\frac{1}{2\kappa}({\bar R}-2\Lambda)-\frac{1}{2}({\bar 
\nabla}\psi)^2\biggl)+\frac{\varepsilon}{\kappa} \int_{\partial{\cal M}} \D^{3}x 
\sqrt{|{\bar h}|}{\bar K},\label{action-Roberts}
\end{align}
which corresponds to the Einstein frame with ${\bar 
V}(\psi)=\Lambda/\kappa$.

In this system, we consider the following topological generalization of 
Roberts-(A)dS solution~\cite{roberts2014,maeda2015}:
\begin{align}
\D s^2=&{\bar g}_{\mu\nu} \D x^\mu \D x^\nu \nonumber \\
=&\biggl(1-\frac{\Lambda}{6}uv\biggl)^{-2}\biggl(-2\D u\D 
v+(-kuv+D_1v^2+D_2u^2){\bar \eta}_{AB}(z)\D \theta^A\D 
\theta^B\biggl),\label{roberts-AdS}
\end{align}
in the coordinates $x^\mu=(u,v,\theta^A)$, where $A, B=2,3$.
In the above solution, $D_1$ and $D_2$ are constants and ${\bar 
\eta}_{AB}$ is the metric on a two-dimensional space of constant 
curvature with its Gauss curvature $k=1,0,-1$.
For $k^2-4D_1D_2>0$, the scalar field $\psi$ is real and given by
\begin{align}
\label{phi-sol1}
\psi=& \left\{
\begin{array}{ll}
\displaystyle{\pm 
\frac{1}{\sqrt{2\kappa}}\ln\biggl|\frac{u\sqrt{k^2-4D_1D_2}+(ku-2D_1v)}{u\sqrt{k^2-4D_1D_2}-(ku-2D_1v)}\biggl|+\psi_0}
& \mbox{for $D_1\ne0$},\\
\displaystyle{\pm 
\frac{1}{\sqrt{2\kappa}}\ln\biggl|D_2-k\frac{v}{u}\biggl|+\psi_1} & 
\mbox{for $D_1=0$},
\end{array} \right.
\end{align}
where $\psi_0$ and $\psi_1$ are constants.
For $k^2-4D_1D_2<0$, $\psi$ is ghost and given by
\begin{align}
\psi=\pm 
i\sqrt{\frac{2}{\kappa}}\left[\arctan\biggl(\frac{ku-2D_1v}{u \sqrt{4D_1D_2-k^2}}\biggl)+\text{sign}(D_1v)\frac{\pi}{2} \right]+\psi_2,
\label{phi-sol2}
\end{align}
where $\psi_2$ is a pure imaginary constant.
If $k^2-4D_1D_2=0$, the field equations give $\psi=$constant and ${\bar 
R}^{\mu\nu}_{~~\rho\sigma}=(\Lambda/3)(\delta^\mu_\rho\delta^\nu_\sigma-\delta^\mu_\sigma\delta^\nu_\rho)$, namely the spacetime is maximally symmetric.
With $\Lambda=0$, the solution (\ref{roberts-AdS}) reduces to the Roberts solution~\cite{roberts1989}.

In~\cite{maeda2009}, it has been presented that a vacuum $C^1$ matching is possible between two Roberts-(A)dS spacetimes with different values of $D_2$ at a null hypersurface $\Sigma$ given by $u=0$.
The induced metric ${\bar h}_{ab}$ on $\Sigma$ is
\begin{eqnarray}
\D s_{\Sigma}^2={\bar h}_{ab}\D y^a \D y^b=D_1v^2{\bar \eta}_{AB}\D \theta^A 
\D \theta^B(={\bar \sigma}_{AB}\D  x^A \D x^B)
\end{eqnarray}
and therefore $D_1\ne 0$ is required, where $y^a=(v,\theta^A)$ is a set of 
coordinates on $\Sigma$.
For $D_1\ne 0$, the value of $\psi$ on $\Sigma$ is constant containing 
$\psi_0$ or $\psi_2$, so we can always set $\psi$ be continuous at 
$\Sigma$ by choosing the value of $\psi_0$ or $\psi_2$ in the spacetime 
attached.

The basis vectors of $\Sigma$ defined by ${\bar e}^\mu_a := \partial 
x^\mu/\partial y^a$ are
\begin{align}
{\bar e}^\mu_v\frac{\partial}{\partial x^\mu}={\bar 
k}^\mu=\frac{\partial}{\partial v},\qquad {\bar 
e}^\mu_A\frac{\partial}{\partial 
x^\mu}=\delta^\mu_{~A}\frac{\partial}{\partial \theta^A},
\end{align}
and the bases are completed by ${\bar N}_\mu \D x^\mu=-\D v$.
They satisfy ${\bar N}_\mu {\bar e}^\mu_v(\equiv {\bar N}_\mu {\bar 
k}^\mu)=-1$ and ${\bar N}_\mu {\bar e}^\mu_A=0$ on $\Sigma$.
Using the following expression
\begin{align}
{\bar\nabla}_\nu {\bar N}_\mu=&\partial_\nu {\bar N}_\mu-{\bar 
\Gamma}^\alpha_{\nu\mu}{\bar N}_\alpha=-{\bar \Gamma}^v_{\nu\mu}{\bar 
N}_v \nonumber\\
=&\frac12{\bar g}^{vu}(\partial_\nu{\bar g}_{\mu u}+\partial_\mu{\bar 
g}_{\nu u}-\partial_u{\bar g}_{\mu\nu}),
\end{align}
and ${\bar g}^{uv}=-(1-\Lambda uv/6)^2$, we compute the non-zero 
components of ${\bar C}_{ab}$ as
\begin{align}
{\bar C}_{vv}=&({\bar\nabla}_\nu {\bar N}_\mu){\bar e}^\mu_{v} {\bar 
e}^\nu_v ={\bar g}^{vu}\partial_v{\bar g}_{vu}=\frac{\Lambda}{3} u\biggl(1-\frac{\Lambda}{6} uv\biggl)^{-1}, \\
{\bar C}_{AB}=&({\bar\nabla}_\nu {\bar N}_\mu){\bar e}^\mu_{A} {\bar 
e}^\nu_B=-\frac12{\bar g}^{vu}\partial_u{\bar g}_{AB} \nonumber \\
=&\frac12\biggl\{\frac{\Lambda}{3}v\biggl(1-\frac{\Lambda}{6}uv\biggl)^{-1}(-kuv+D_1v^2+D_2u^2)+(-kv+2D_2u)\biggl\}{\bar
\eta}_{AB},
\end{align}
and hence
\begin{align}
{\bar C}_{vv}|_\Sigma=0, \qquad {\bar 
C}_{AB}|_\Sigma=\frac12v\biggl(\frac{1}{3}\Lambda D_1v^2-k\biggl){\bar 
\eta}_{AB}.
\end{align}
Since ${\bar h}_{ab}$ and ${\bar C}_{ab}|_\Sigma$ do not contain $D_2$, 
two Roberts-(A)dS spacetimes (\ref{roberts-AdS}) with the same nonzero 
$D_1$ but different $D_2$ can be attached at $u=0$ in a $C^1$ regular manner, where $[{\bar h}_{ab}]=[{\bar C}_{ab}]=0$ are realized.
Then, by Proposition~\ref{Prop:C1matching-null-E}, ${\bar t}_{\mu\nu}=0$ holds and hence there is no massive thin-shell at $\Sigma$.
As a special case, a Roberts-(A)dS spacetime can be attached to the past 
(A)dS spacetime at $u=0$ and the resulting spacetime may represent 
black-hole or naked-singularity formation from a regular initial datum.

Lastly, let us see whether ${\bar W}:=-{\bar N}^\mu[\partial_\mu\psi]$ is 
vanishing or not at $u=0$.
With the following expression;
\begin{align}
{\bar N}^\mu\frac{\partial}{\partial x^\mu}=\biggl(1-\frac16\Lambda 
uv\biggl)^2\frac{\partial}{\partial u},
\end{align}
we obtain
\begin{align}
({\bar N}^\rho\partial_\rho \psi)|_\Sigma=\pm 
\frac{\sqrt{k^2-4D_1D_2}}{\sqrt{2\kappa}D_1v}
\end{align}
for $k^2-4D_1D_2\ne 0$ with $D_1\ne0$.
Since the above expression contains both $D_1$ and $D_2$, ${\bar W}\ne 0$ holds when two Roberts-(A)dS spacetimes (\ref{roberts-AdS}) with the same nonzero $D_1$ but different $D_2$ are attached at $u=0$.
Therefore, Proposition~\ref{Prop:null-regularity-relation} does not work and the $C^1$ regularity at $\Sigma$ in the Jordan frame is not clear in this case.
However, since ${\bar t}_{\mu\nu}=0$ holds at $\Sigma$, $t_{\mu\nu}=0$ also holds in the Jordan frame under the assumptions in Proposition~\ref{Prop:null}.

\subsubsection{Generalized Xu solution in the Jordan frame ($n=3$)}
Another example is presented in the three-dimensional gravity coupled to 
a non-minimally self-interacting scalar field $\phi$ in the presence of 
a negative cosmological constant $\Lambda$:
\begin{align}
I_{\rm J}=&\int \D^{3}x \sqrt{-g}\left( 
\frac{1}{2\kappa}(R -2 \Lambda)-\frac{1}{2}(\nabla\phi)^2-\frac{1}{16}R 
\phi^2-\alpha \phi^6 \right) \nonumber \\
&+ \frac{\varepsilon}{\kappa}\int_{\partial{\cal M}} \D^{2}x 
\sqrt{|h|}\biggl(1-\frac{\kappa}{8}\phi^2\biggl)K,  \label{action-1000}
\end{align}
which is the Jordan frame with
\begin{align}
f(\phi)=&\frac{1}{2\kappa}\biggl(1-\frac{\kappa}{8}\phi^2\biggl),\label{form-f}\\
V(\phi)=&-\frac{1}{\kappa l^2}+\alpha \phi^6,
\end{align}
where $l$ is the AdS radius defined by $l^{-2}:=-\Lambda$.
To simplify the expressions in the following argument, we introduce a 
constant $\beta$ defined by
\begin{equation}
\beta:= \frac{512 \alpha l^2-\kappa^2}{8 \kappa l^2}. \label{def-beta}
\end{equation}

In this system, there is the following generalized Xu solution~\cite{Xu:2014xqa,Aviles:2018vnf}:
\begin{align}
\D s^2=&-f(v,r)\D v^2+2\D v\D r+r^2\D\theta^2,\label{sol-attach} \\
\phi(v,r)=&\frac{a(v)}{\sqrt{r+\kappa a(v)^2/8}},\label{def-phi0}
\end{align}
where the metric function $f(v,r)$ is given by
\begin{align}
f(v,r)=&\frac{r^2}{l^2}-B_0 a(v)-\frac{B_0 \kappa  a(v)^3}{12r}.
\label{def-f0}
\end{align}
Here $B_0$ is a parameter in the solution and the function $a(v)$ is 
given by
\begin{eqnarray}
a(v)=\frac{2B_0 }{3 \kappa }v \label{param-beta=0}
\end{eqnarray}
for $\beta=0$ and
\begin{equation} \label{param}
\frac12\ln \left(\frac{(a-a_0)^2}{a^2+a_0 a+a_0^2}\right)-\sqrt{3} 
\arctan\left(\frac{2 a+a_0}{\sqrt{3} a_0}\right)=\frac{3 a_0^2 \beta 
}{4} \biggl(v-\frac{2\sqrt{3}\pi   }{ 9 \beta a_0^2}\biggl)
\end{equation}
for $\beta\ne 0$.
Here $B_0$ is an integration constant and $a_0$ is defined by
\begin{equation}
a_0:=-\epsilon\biggl|\frac{8B_0 }{3 \kappa 
\beta}\biggl|^{1/3},\label{a0-def}
\end{equation}
where $\epsilon$ is the sign of $8B_0/(3 \kappa \beta)$, and hence we have $B_0=-(3/8)\kappa \beta a_0^3$.
In the solutions (\ref{param-beta=0}) and (\ref{param}), we have set 
another integration constant such that $a(0)=0$ without loss of generality.
$B_0=0$ gives the massless BTZ spacetime and the behavior of $a$ near 
$v=0$ for $\beta\ne 0$ is given by
\begin{equation}
a(v)\simeq \frac{2B_0}{3\kappa}v, \label{asymp-a0}
\end{equation}
which is the same as Eq.~(\ref{param-beta=0}) for $\beta=0$.

Actually, the generalized Xu solution (\ref{sol-attach}) for $v\ge 0$ 
can be attached to the massless BTZ spacetime for $v\le 0$ with 
$\phi\equiv 0$ (and hence $[\phi]=0$ is realized).
On the null hypersurface $\Sigma$ defined by $v=0$, we install 
coordinates $y^a=({\lambda},\theta^A)$ which are the same on both past 
and future sides of $\Sigma$.
Here ${\lambda}$ is an arbitrary parameter on the null generators of 
$\Sigma$ and $\theta^A$ label the generators, where the index $A$ is 
always $A=1$ in the three-dimensional case.
We identify $-r$ with ${\lambda}$ and set $\theta^A=\theta$ on $\Sigma$ 
in the spacetime (\ref{sol-attach}).

The parametric equations $x^\mu=x^\mu({\lambda},\theta^A)$ describing 
$\Sigma$ are $v=0$, $r=-{\lambda}$, and $\theta=\theta$.
The line element on $\Sigma$ is one-dimensional and given by
\begin{align}
\D s_{\Sigma}^2=h_{ab}\D y^a \D y^b={\lambda}^2\D\theta^2(=\sigma_{AB}\D \theta^A \D 
\theta^B),\label{hab}
\end{align}
where $y^a=(\lambda,\theta)$ is a set of coordinates on $\Sigma$.
Using them, we obtain the tangent vectors of $\Sigma$ defined by 
$e^\mu_a := \partial x^\mu/\partial y^a$ as
\begin{align}
e^\mu_{\lambda}\frac{\partial}{\partial x^\mu}=-\frac{\partial}{\partial 
r},\qquad e^\mu_\theta\frac{\partial}{\partial 
x^\mu}=\frac{\partial}{\partial \theta}.
\end{align}
An auxiliary null vector $N^\mu$ given by
\begin{align}
N^\mu \frac{\partial}{\partial x^\mu}=\frac{\partial}{\partial 
v}+\frac12f(0,r)\frac{\partial}{\partial r} \label{N-attachment}
\end{align}
completes the basis.
The expression $N_\mu \D x^\mu=-(f(0,r)/2)\D v+\D r$ shows $N_\mu N^\mu 
=0$, $N_\mu e^\mu_{\lambda}=-1$, and $N_\mu e^\mu_\theta=0$.
Then, the only nonvanishing component of the transverse curvature 
$C_{ab}:=(\nabla_\nu N_{\mu}) e^\mu_{a} e^\nu_b$ of $\Sigma$ is
\begin{align}
C_{\theta\theta}=\frac12rf(0,r)=\frac{r}{2l^2}.\label{trans-C0}
\end{align}
Since Eqs.~(\ref{hab}) and (\ref{trans-C0}) do not contain $B_0$, 
$[\sigma_{AB}]=[C_{ab}]=[\phi]=0$ is realized and therefore a $C^1$ regular matching is achieved at $\Sigma$.

On the other hand, $W\ne 0$ holds at $v=0$ because 
Eqs.~(\ref{N-attachment}) and (\ref{def-phi0}) show
\begin{align}
(N^\mu\nabla_\mu\phi)|_{\Sigma}=\frac{2 B_0}{3 \kappa r^{1/2}}, 
\label{phi-jump}
\end{align}
which contains $B_0$.
Nevertheless, since Eq.~(\ref{form-f}) shows $f'(\phi_\Sigma)=0$ with $\phi_\Sigma=0$, the J-vacuum condition (III) in 
Proposition~\ref{Prop:C1-null-J} is satisfied at $v=0$.
In this case, since $f(\phi_\Sigma)=1/2\kappa \ne 0$ holds, both $C^1$ regularity and vacuum $\Sigma$ are realized in the Einstein frame by Propositions~\ref{Prop:null-regularity-relation} and \ref{Prop:null}.

\section{Summary}
\label{sec:conclusion}

In the present paper, we have studied junction conditions in a large class of scalar-tensor theories in arbitrary $n(\ge 3)$ dimensions.  
We have treated both null and non-null junction hypersurfaces $\Sigma$ under the assumptions (A) the bulk energy-momentum tensor does not contribute to the energy-momentum tensor on $\Sigma$ and (B) the energy-momentum tensor on $\Sigma$ does not contain the same scalar field $\phi$ in the bulk spacetime.
While the metric and scalar field must be continuous on $\Sigma$ as the first junction conditions, the jumps of their first derivatives and the matter field on $\Sigma$ are related as the second junction conditions given from the Einstein equations and the equation of motion for $\phi$ treated as distributions~\cite{Poisson:2002nv}.
In the case of non-null $\Sigma$, the resulting junction conditions agree with the ones obtained in the variational method demonstrated in~\cite{cr1999}. 
 
Subsequently, we have clarified the $C^1$ regular matching conditions and the vacuum conditions at $\Sigma$ both in the Jordan and Einstein frames.
At non-null $\Sigma$ in the Einstein frame, the $C^1$ regularity (E-regularity) is equivalent to the vacuum $\Sigma$ condition (E-vacuum).
In the Jordan frame, in contrast, while the $C^1$ regularity (J-regularity) implies vacuum $\Sigma$ (J-vacuum), J-vacuum does not necessarily imply J-regularity.
In other words, J-regularity is a sufficient condition for J-vacuum which suggests a possibility of {\it vacuum} thin-shell at non-null $\Sigma$ in the Jordan frame.

The situations are different in the case where $\Sigma$ is null.
In this case, E-regularity and E-vacuum are even not equivalent.
Actually, E-regularity is a sufficient condition for E-vacuum so that there is a possibility of vacuum thin-shell at null $\Sigma$.
To compound matters, J-regularity and J-vacuum do not necessarily imply each other, which suggests that both non-vacuum $C^1$ regular matching and vacuum thin-shell may be possible at null $\Sigma$ in the Jordan frame.

Lastly, we have clarified the relations between the sufficient conditions for the $C^1$ regularity in the Jordan and Einstein frames and also between the vacuum $\Sigma$ conditions, which allow us to identify the properties of the junction hypersurface $\Sigma$ in the other frame.
We have adopted these results to two concrete exact solutions; The Roberts-(A)dS solution in the Einstein frame in four dimensions and the generalized Xu solution in the Jordan frame in three dimensions.

As demonstrated in these two examples, all the results in the present paper may provide a firm basis for applications in a variety of contexts, which would clarify the effects of the non-minimal coupling of the scalar field to gravity.
Additionally, to construct concrete examples of the following configurations is an interesting task: (i) a vacuum thin-shell at null $\Sigma$ in the Einstein frame, (ii) a vacuum thin-shell at null and non-null $\Sigma$ in the Jordan frame, and (iii) a non-vacuum $C^1$ regular matching at null $\Sigma$ in the Jordan frame.
We leave these problems for future investigations.

\noindent{\bf{Acknowledgements}}\\[3mm]
CM~thanks Hokkai-Gakuen University for a kind hospitality, where a part of this work was carried out. 
This work has been partially funded by the Fondecyt
grants  1161311 and 1180368. LA thanks the Conicyt grant 21160827. The Centro de Estudios Cient\'{\i}ficos (CECs) is funded by the Chilean Government through the Centers of Excellence Base Financing Program of Conicyt.

\appendix

\section{Transformation from Jordan to Einstein frame}
\label{App:JtoE}
In this appendix, we consider a conformal transformation from the following action in the Jordan frame:
\begin{align} 
I_{\rm J}=&\int_{\cal M} \D^{n}x \sqrt{-g}\biggl( f(\phi)R-\frac{1}{2}(\nabla\phi)^2 -V(\phi) \biggl)+2\varepsilon\int_{\Sigma} \D^{n-1}y \sqrt{|h|}f(\phi)K \nonumber \\
&+\int_{\cal M} \D^{n}x \sqrt{-g}{\cal L}_{\cal M}^{(m)}+\int_{\Sigma} \D^{n-1}x \sqrt{|h|}{\cal L}_{\Sigma}^{(m)}.\label{J-action-appendix}
\end{align}

By a conformal transformation $g_{\mu\nu}(x)=\Omega(x)^2{\bar 
g}_{\mu\nu}$ in $n$ dimensions, the Ricci scalar is transformed as
\begin{eqnarray}
R&=&\Omega^{-2}\left\{{\bar R}-2(n-1){\bar 
\Box}\ln\Omega-(n-1)(n-2)({\bar \nabla}_\rho\ln\Omega)({\bar 
\nabla}^\rho\ln\Omega)\right\},
\end{eqnarray}
which is shown by the following transformation of the Christoffel symbol:
\begin{align}
{\Gamma}^\mu_{\rho\sigma}=&\frac12g^{\mu\alpha}(\partial_\sigma 
g_{\alpha\rho}+\partial_\rho g_{\alpha\sigma}-\partial_\alpha 
g_{\rho\sigma}) \nonumber \\
=&\frac12\Omega^{-2}{\bar g}^{\mu\alpha}\biggl\{\partial_\sigma 
(\Omega^2{\bar g}_{\alpha\rho})+\partial_\rho (\Omega^2{\bar 
g}_{\alpha\sigma})-\partial_\alpha (\Omega^2{\bar 
g}_{\rho\sigma})\biggl\} \nonumber \\
=&{\bar \Gamma}^\mu_{\rho\sigma}+(\partial_\sigma\ln\Omega){\bar 
g}^{\mu}_{~\rho}+(\partial_\rho \ln\Omega){\bar 
g}^{\mu}_{~\sigma}-(\partial_\alpha \ln\Omega){\bar g}^{\mu\alpha}{\bar 
g}_{\rho\sigma}.\label{Gamma-trans}
\end{align}
We consider the case where the matching non-null hypersurface $\Sigma$ is 
described by $\Phi(x)=0$ in both frames.
In this case, the unit orthogonal vector of $\Sigma$ is transformed as
\begin{align}
{\bar n}_\mu:=&(\varepsilon {\bar g}^{\rho\sigma}\nabla_\rho 
\Phi\nabla_\sigma \Phi)^{-1/2}\nabla_\mu \Phi = \Omega^{-1}(\varepsilon 
g^{\rho\sigma}\nabla_\rho \Phi\nabla_\sigma \Phi)^{-1/2}\nabla_\mu \Phi 
=\Omega^{-1} n_\mu\label{n-def-E}
\end{align}
and hence the projection tensor is transformed as
\begin{align}
{\bar h}_{\mu\nu}:={\bar g}_{\mu\nu}-\varepsilon {\bar n}_\mu {\bar 
n}_\nu=\Omega^{-2}(g_{\mu\nu}-\varepsilon n_\mu n_\nu)= 
\Omega^{-2}h_{\mu\nu}.
\end{align}
Using these results, one can show that the extrinsic curvature and its 
trace are transformed as
\begin{align}
K_{\mu\nu}=&h^{~\rho}_{(\mu} h^{~\sigma}_{\nu)}\nabla_\rho n_{\sigma} 
={\bar h}^{~\rho}_{(\mu} {\bar h}^{~\sigma}_{\nu)}(\partial_\rho 
n_{\sigma}-\Gamma^\alpha_{\rho\sigma}n_\alpha)  \nonumber \\
=&{\bar h}^{~\rho}_{(\mu} {\bar 
h}^{~\sigma}_{\nu)}\biggl\{\partial_\rho(\Omega {\bar n}_{\sigma}) 
-\biggl({\bar 
\Gamma}^\alpha_{\rho\sigma}+(\partial_\sigma\ln\Omega){\bar 
g}^{\alpha}_{~\rho}+(\partial_\rho \ln\Omega){\bar 
g}^{\alpha}_{~\sigma}-(\partial_\beta \ln\Omega){\bar 
g}^{\alpha\beta}{\bar g}_{\rho\sigma}\biggl)\Omega {\bar 
n}_\alpha\biggl\} \nonumber \\
=&\Omega{\bar h}^{~\rho}_{(\mu} {\bar 
h}^{~\sigma}_{\nu)}\left\{{\bar\nabla}_\rho {\bar 
n}_{\sigma}+(\partial_\beta \ln\Omega){\bar g}^{\alpha\beta}{\bar 
g}_{\rho\sigma}{\bar n}_\alpha\right\}=\Omega\left\{{\bar 
K}_{\mu\nu}+{\bar h}_{\mu\nu}(\partial_\beta \ln\Omega){\bar 
n}^\beta\right\} \label{rel-KK}
\end{align}
and
\begin{align}
{K}=&g^{\mu\nu}K_{\mu\nu}=\Omega^{-1}{\bar g}^{\mu\nu}\left\{{\bar 
K}_{\mu\nu}+{\bar h}_{\mu\nu}(\partial_\sigma \ln\Omega){\bar 
n}^\sigma\right\} \nonumber \\
=&\Omega^{-1}{\bar K}+(n-1)\Omega^{-1}(\partial_\sigma\ln\Omega){\bar 
n}^\sigma.
\end{align}

Putting the above expressions into the action (\ref{J-action-appendix}) in the 
Jordan frame, we obtain
\begin{align}
I_{\rm J}=&\int_{\cal M} \D^{n}x \sqrt{-{\bar g}}\biggl\{ 
\Omega^{n-2}f(\phi)\biggl({\bar R}-2(n-1){\bar 
\Box}\ln\Omega-(n-1)(n-2)({\bar \nabla}_\rho\ln\Omega)({\bar 
\nabla}^\rho\ln\Omega)\biggl) \nonumber \\
&-\frac{1}{2}\Omega^{n-2}({\bar \nabla}\phi)^2 -\Omega^nV(\phi) \biggl\} 
+2\varepsilon \int_{\Sigma} \D^{n-1}x \sqrt{|{\bar 
h}|}\Omega^{n-2}f(\phi)\biggl({\bar K}+(n-1)({\bar 
\nabla}_\sigma\ln\Omega){\bar n}^\sigma\biggl) \nonumber \\
&+\int_{\cal M} \D^{n}x \sqrt{-{\bar g}}\Omega^n{\cal L}_{\cal 
M}^{(m)}+\int_{\Sigma} \D^{n-1}x \sqrt{|{\bar h}|}\Omega^{n-1}{\cal 
L}_{\Sigma}^{(m)} \nonumber \\
=&\int_{\cal M} \D^{n}x \sqrt{-{\bar g}}\biggl\{ 
\Omega^{n-2}f(\phi)\biggl({\bar R}-(n-1)(n-2)({\bar 
\nabla}_\rho\ln\Omega)({\bar \nabla}^\rho\ln\Omega)\biggl) \nonumber \\
&+2(n-1){\bar\nabla}^\rho(\Omega^{n-2}f(\phi)){\bar\nabla}_\rho\ln\Omega-\frac{1}{2}\Omega^{n-2}({\bar
\nabla}\phi)^2 -\Omega^nV(\phi) \biggl\}  \nonumber \\
&+2\varepsilon \int_{\Sigma} \D^{n-1}x \sqrt{|{\bar 
h}|}\Omega^{n-2}f(\phi){\bar K}+\int_{\cal M} \D^{n}x \sqrt{-{\bar 
g}}\Omega^n{\cal L}_{\cal M}^{(m)}+\int_{\Sigma} \D^{n-1}x \sqrt{|{\bar 
h}|}\Omega^{n-1}{\cal L}_{\Sigma}^{(m)},
\end{align}
where we used the Stokes' theorem (\ref{stokes}) at the second equality.

Setting $\Omega=(2\kappa_n f(\phi))^{-1/(n-2)}$, we obtain the action in 
the Einstein frame:
\begin{align}
I_{\rm E}=&\int_{\cal M} \D^{n}x \sqrt{-{\bar g}}\biggl\{ 
\frac{1}{2\kappa_n}{\bar 
R}-\frac{2(n-1){f'}^2+(n-2)f}{4(n-2)\kappa_nf^2}({\bar \nabla}\phi)^2 
-\Omega^nV(\phi) \biggl\}  \nonumber \\
&+\frac{\varepsilon}{\kappa_n} \int_{\Sigma} \D^{n-1}x \sqrt{|{\bar 
h}|}{\bar K}+\int_{\cal M} \D^{n}x \sqrt{-{\bar g}}\Omega^n{\cal 
L}_{\cal M}^{(m)}+\int_{\Sigma} \D^{n-1}x \sqrt{|{\bar 
h}|}\Omega^{n-1}{\cal L}_{\Sigma}^{(m)}.
\end{align}
By a redefinition of the scalar field (\ref{def-psi}), we finally write down the action in the Einstein frame in the following 
canonical form:
\begin{align}
I_{\rm E}=&\int_{\cal M} \D^{n}x \sqrt{-{\bar g}}\biggl\{ 
\frac{1}{2\kappa_n}{\bar R}-\frac12({\bar \nabla}\psi)^2 
-{\bar V}(\psi) \biggl\}  \nonumber \\
&+\frac{\varepsilon}{\kappa_n} \int_{\Sigma} \D^{n-1}x \sqrt{|{\bar 
h}|}{\bar K}+\int_{\cal M} \D^{n}x \sqrt{-{\bar g}}{\bar {\cal L}}_{\cal M}^{(m)}+\int_{\Sigma} \D^{n-1}x \sqrt{|{\bar 
h}|}{\bar {\cal L}}_{\Sigma}^{(m)},\label{conformal-last}
\end{align}
where
\begin{align}
{\bar V}(\psi):=&(2\kappa_n 
f(\phi(\psi)))^{-n/(n-2)}V(\phi(\psi)),\label{Vbar-V}\\
{\bar {\cal L}}_{\cal M}^{(m)}:=&(2\kappa_n 
f(\phi(\psi)))^{-n/(n-2)}{\cal L}_{\cal M}^{(m)},\label{TT-M}\\
{\bar {\cal L}}_{\Sigma}^{(m)}:=&(2\kappa_n 
f(\phi(\psi)))^{-(n-1)/(n-2)}{\cal L}_{\Sigma}^{(m)}. \label{tt-Sigma}
\end{align}

As explained in Sec.~\ref{sec:proper}, for a proper mapping between the bulk equations in the Jordan and Einstein frames, assumptions in Lemma~\ref{lm:correspondence-basic} are required.
For a proper mapping between the junction conditions in two frames, one needs $\sqrt{-g}{\cal L}_{{\cal M}}^{(m)}= \sqrt{-{\bar g}}{\bar {\cal L}}_{{\cal M}}^{(m)}$ in addition, which includes the vacuum case ${\cal L}_{\Sigma}^{(m)}={\bar {\cal L}}_{\Sigma}^{(m)}\equiv 0$.

\section{Junction conditions from variational principle for non-null $\Sigma$}
\label{app:variation-nonnull}
In this appendix, we derive the junction conditions in the Jordan frame by the variational principle in the case where the matching hypersurface $\Sigma$ is non-null. For this purpose it is convenient to start with the following action:
\begin{equation} 
I_0=I_{\cal M}+I_{\partial{\cal M}},\label{total-action-app}
\end{equation}
where the bulk  ($I_{\cal M}$) and boundary ($I_{\partial{\cal M}}$) actions are given by 
\begin{align} 
I_{\cal M}:=&\int_{\cal M} \D^{n}x \sqrt{-g}\biggl( f(\phi)R-\frac{1}{2}g^{\mu \nu}(\nabla_\mu \phi)( \nabla_\nu \phi) -V(\phi)+{\cal L}^{(m)}\biggl),\label{action-bt2-app}\\
I_{\partial{\cal M}}:=&2\varepsilon \int_{\partial{\cal M}}\D^{n-1}y\sqrt{|h|}f(\phi)K.\label{action-bt3-app}
\end{align}

In general relativity ($f(\phi)=1/2\kappa_n$), Eq.~(\ref{action-bt3-app}) reduces to the Gibbons-Hawking term \cite{ Gibbons-Hawking}. Such a boundary term has been constructed also in Einstein-Gauss-Bonnet gravity~\cite{davis2003,gw2003} and further generalized in Lovelock gravity~\cite{mo2007}, which is the most general quasi-linear second-order theory of gravity in arbitrary dimensions~\cite{Lovelock}.

In the following, we assume that $g_{\mu\nu}$ and $\phi$ are continuous at $\partial{\cal M}$ and matter Lagrangian density $\sqrt{-g}{\cal L}^{(m)}_{\cal M}$  does not depend on $\phi$.

\subsection{Useful formulae}
For variation, we will use
\begin{align}
\delta g^{\mu\alpha}=-g^{\nu\alpha}g^{\mu\rho}\delta g_{\nu\rho},\qquad \delta g_{\nu\alpha}=-g_{\mu\alpha} g_{\nu\rho}\delta g^{\mu\rho}.  \label{lemma2}
\end{align}
Jacobi's formula shows  
\begin{align}
\delta\sqrt{-g}=-\frac{1}{2}\sqrt{-g}g_{\mu\nu}\delta g^{\mu \nu} \label{lemma3}
\end{align}
and
\begin{align}
\delta\sqrt{|h|}=&-\frac{1}{2}\sqrt{|h|}h_{ab}\delta h^{ab}=-\frac{1}{2}\sqrt{|h|}h_{\mu\nu}\delta h^{\mu\nu}=-\frac{1}{2}\sqrt{|h|}h_{\mu\nu}\delta g^{\mu\nu}.\label{lemma4}
\end{align}

While $\Gamma^\rho_{\mu\nu}$ is not a tensor, its variation $\delta\Gamma^\rho_{\mu\nu}$ is a tensor given by 
\begin{align}
\delta\Gamma^\rho_{\mu\nu}=\frac12g^{\rho\alpha}(\nabla_\nu \delta g_{\alpha\mu}+\nabla_\mu \delta g_{\alpha\nu}-\nabla_\alpha \delta g_{\mu\nu}),\label{delta-Gamma}
\end{align}
which gives
\begin{align}
\delta R^\rho_{~\sigma\mu\nu}=&\nabla_\mu\delta\Gamma^\rho_{\nu\sigma}-\nabla_\nu\delta\Gamma^\rho_{\mu\sigma}, \\
\delta R_{\sigma\nu}=&\nabla_\rho\delta\Gamma^\rho_{\nu\sigma}-\nabla_\nu\delta\Gamma^\rho_{\rho\sigma}.\label{lemma5}
\end{align}
We can rewrite the term $f(\phi)g^{\mu\nu}\delta R_{\mu\nu}$ such that 
\begin{align} 
f(\phi)g^{\mu\nu}\delta R_{\mu\nu}=&\nabla_\rho J^\rho-(\nabla_\mu\nabla_\rho f(\phi))\delta g^{\rho\mu} +(\Box f(\phi))g_{\mu\nu}\delta g^{\mu\nu},
\end{align}
where
\begin{align}
J^\rho:=&f(\phi)\biggl(-\nabla_\mu\delta g^{\rho\mu}+\nabla_\alpha(g_{\mu\nu}g^{\rho\alpha} \delta g^{\mu\nu})\biggl)+(\nabla_\mu f(\phi))\delta g^{\rho\mu}-(\nabla_\alpha f(\phi)) g_{\mu\nu}g^{\rho\alpha} \delta g^{\mu\nu}. \label{def-J}
\end{align}

From Eq.~(\ref{n-def}), we obtain
\begin{align}
\delta n_\mu=&\frac12\varepsilon  n_\mu n^\alpha n^\beta \delta g_{\alpha\beta},\\
\delta n^\mu=&-g^{\mu\alpha}n^\beta\delta g_{\alpha\beta}+\frac12\varepsilon  n^\mu n^\alpha n^\beta \delta g_{\alpha\beta}.
\end{align}
Using this, after lengthy but straightforward calculations, we obtain
\begin{align}
\delta K=&-\frac12(\nabla^{\alpha}n^\beta)\delta g_{\alpha\beta} +\frac12\varepsilon n^{\alpha}n^{\mu}g^{\beta\nu}(\nabla_{\mu}n_\nu)\delta g_{\alpha\beta} \nonumber \\
&-\frac12n^\beta h^{\alpha\mu}(\nabla_\mu \delta g_{\alpha\beta}-\nabla_\beta \delta g_{\alpha\mu})-\frac12h^\rho_\mu\nabla_\rho (h^{\mu\alpha} n^\beta \delta g_{\alpha\beta}).
\end{align}


\subsection{Variation with respect to $\phi$}
First let us consider variation with respect to $\phi$.
Using integration by parts and used the Stokes' theorem (\ref{stokes}), variation of the bulk action (\ref{action-bt2-app}) leads
\begin{align} 
\delta_\phi I_{\cal M}=\int_{\cal M} \D^{n}x \sqrt{-g}\biggl(\Box\phi+f'(\phi)R-V'(\phi)\biggl)\delta \phi-\int_{\partial{\cal M}}\D^{n-1}y\sqrt{|h|}(\varepsilon n^\mu\nabla_\mu\phi)\delta\phi.\label{action-bt-phi2}
\end{align}
On the other hand, variation of the boundary term (\ref{action-bt3-app}) simply leads
\begin{align}
\delta_\phi I_{\partial{\cal M}}=2\varepsilon \int_{\partial{\cal M}}\D^{n-1}y\sqrt{|h|}f'(\phi)K\delta\phi.
\end{align}

Since matter Lagrangian density $\sqrt{-g}{\cal L}^{(m)}_{\cal M}$ does not depend on $\phi$, variation of the total action (\ref{total-action-app}) with respect to $\phi$ reduces to the following form:
\begin{align} 
\delta_\phi I_0=\int_{\cal M}\D^{n}x \sqrt{-g}{\cal E}_{(\phi)}\delta\phi +\int_{\partial{\cal M}}\D^{n-1}y\sqrt{|h|}{\cal F}_{(\phi)}\delta\phi,
\end{align}
where
\begin{align} 
{\cal E}_{(\phi)}:=&\Box\phi +f'(\phi)R-V'(\phi),\\
{\cal F}_{(\phi)}:=&\varepsilon \left(2f'(\phi)K-n^\mu\nabla_\mu\phi\right).
\end{align}

\subsection{Variation with respect to $g^{\mu\nu}$}
Next let us consider variation with respect to $g^{\mu\nu}$.
Using integration by parts and the Stokes' theorem (\ref{stokes}), variation of the bulk action (\ref{action-bt2-app}) leads
\begin{align} 
\delta_g I_{\cal M}=&\int_{\cal M} \D^{n}x\sqrt{-g}\biggl({\cal E}_{\mu\nu}-\frac12T_{\mu\nu}\biggl)\delta g^{\mu\nu}+\varepsilon \int_{\partial{\cal M}} \D^{n-1}y \sqrt{|h|}n_\rho {J}^\rho \nonumber \\
=&\int_{\cal M} \D^{n}x\sqrt{-g}\biggl({\cal E}_{\mu\nu}-\frac12T_{\mu\nu}\biggl)\delta g^{\mu\nu} \nonumber \\
&+\varepsilon \int_{\partial{\cal M}} \D^{n-1}y \sqrt{|h|}\biggl\{f(\phi)n^\sigma h^{\mu\nu}\biggl(\nabla_\mu(\delta g_{\sigma\nu})-\nabla_\sigma(\delta g_{\mu\nu})\biggl)\nonumber \\
&~~~~~~~~~~~~~~~~~~~~~~~~~~~~~~ -n^\sigma(\nabla_\mu f(\phi))g^{\mu\nu}\delta g_{\sigma\nu}+n^\sigma(\nabla_\sigma f(\phi))g^{\mu\nu}\delta g_{\mu\nu}\biggl\},\label{deltaIM-g}
\end{align}
where ${J}^\rho$ is defined by Eq.~(\ref{def-J}) and 
\begin{align} 
{\cal E}_{\mu\nu}:=&f(\phi)G_{\mu\nu}+\frac12g_{\mu\nu}\biggl(\frac{1}{2}(\nabla\phi)^2+V(\phi) \biggl) \nonumber \\
&-\frac12(\nabla_\mu \phi)(\nabla_\nu \phi) -\nabla_\mu\nabla_\nu f(\phi) +g_{\mu\nu}\Box f(\phi),\label{calE-g}
\end{align}
with
\begin{align} 
T_{\mu\nu}:=-2\frac{\partial {\cal L}^{(m)}_{\cal M}}{\partial g^{\mu\nu}}+g_{\mu\nu}{\cal L}^{(m)}_{\cal M}.
\end{align} 
On the other hand, variation of the boundary term (\ref{action-bt3-app}) leads
\begin{align}
\delta_gI_{\partial{\cal M}}=&\varepsilon \int_{\partial{\cal M}}\D^{n-1}y\sqrt{|h|}\biggl[f(\phi)\biggl\{(K h^{\alpha\beta}-K^{\alpha\beta})\delta g_{\alpha\beta} -n^\beta h^{\alpha\mu}(\nabla_\mu \delta g_{\alpha\beta}-\nabla_\beta \delta g_{\alpha\mu})\biggl\}\nonumber \\
&+(\nabla_\rho f(\phi))h^{\rho\alpha} n^\beta \delta g_{\alpha\beta}-h^\rho_\mu\nabla_\rho(f(\phi) h^{\mu\alpha} n^\beta \delta g_{\alpha\beta})\biggl].
\end{align}
The last term in the above integrand is a total derivative term on $\partial{\cal M}$ and becomes a surface integral at $\partial \partial{\cal M}$, namely the boundary of $\partial{\cal M}$ because for a given vector $v^\mu$, we have 
\begin{align} 
h_\mu^{~\rho}\nabla_\rho v^\mu=&h^{\mu\rho}\nabla_\rho v_\mu=e^\mu_a e^\rho_b h^{ab}\nabla_\rho v_\mu=h^{ab}D_b v_a=D_a v^a.
\end{align}
Hence, we have
\begin{align} 
\delta_g (I_{\cal M}+I_{\partial{\cal M}})=&\int_{\cal M} \D^{n}x\sqrt{-g}\biggl({\cal E}_{\mu\nu}-\frac12T_{\mu\nu}\biggl)\delta g^{\mu\nu} \nonumber \\
&-\varepsilon \int_{\partial{\cal M}} \D^{n-1}y \sqrt{|h|}\biggl\{f(\phi)(K h_{\mu\nu}-K_{\mu\nu})+n^\sigma(\nabla_\sigma f(\phi))h_{\mu\nu}\biggl\}\delta g^{\mu\nu} \nonumber \\
&+(\mbox{surface integral at $\partial \partial{\cal M}$}),
\end{align}
where we used Eq.~(\ref{lemma2}).

Now we have shown that variation of the total action (\ref{total-action-app}) with respect to $g^{\mu\nu}$ reduces to the following form:
\begin{align} 
\delta_g I_0=&\int_{\cal M} \D^{n}x\sqrt{-g}\biggl({\cal E}_{\mu\nu}-\frac12T_{\mu\nu}\biggl)\delta g^{\mu\nu}-\int_{\partial{\cal M}} \D^{n-1}y \sqrt{|h|}{\cal F}_{\mu\nu}\delta g^{\mu\nu}\nonumber \\
&+(\mbox{Surface integral at $\partial\partial{\cal M}$}),\label{deltaI-g-last}
\end{align}
where  
\begin{align} 
{\cal F}_{\mu\nu}:=\varepsilon f(\phi)(K h_{\mu\nu}-K_{\mu\nu})+\varepsilon n^\sigma(\nabla_\sigma f(\phi))h_{\mu\nu}.
\end{align}

\subsection{Derivation of the junction conditions}

Assume that the spacetime ${\cal M}$ consists of two parts ${\cal M_+}$ and ${\cal M_-}$. In a situation where ${\cal M_+}$ and ${\cal M_-}$ are connected at a non-null hypersurface $\Sigma$ as described in Fig.~\ref{Fig-Variation-hypersurface}, we propose the following action 
\begin{align}
I_{\rm J}=&\int_{\cal M_+} \D^{n}x_+ \sqrt{-g^+}\biggl(f(\phi^+)R^+-\frac{1}{2}(\nabla\phi^+)^2 -V(\phi^+)+{\cal L}_{\cal M_+}^{(m)} \biggl) \nonumber \\
&+\int_{\cal M_-} \D^{n}x_- \sqrt{-g^-}\biggl(f(\phi^-)R^--\frac{1}{2}(\nabla\phi^-)^2 -V(\phi^-) +{\cal L}_{\cal M_-}^{(m)}\biggl) \nonumber \\
&+2\epsilon_+ \int_{\partial{\cal M}_+-\Sigma_{+}} \D^{n-1}z_+ \sqrt{|\zeta^+|}f(\phi^+)K^++2\epsilon_- \int_{\partial{\cal M}_--\Sigma_{-}} \D^{n-1}z_- \sqrt{|\zeta^-|}f(\phi^-)K^- \nonumber \\
&+2\varepsilon \int_{\Sigma_{+}} \D^{n-1}y \sqrt{|h|}f(\phi)K^+ +2\varepsilon \int_{\Sigma_{-}} \D^{n-1}y \sqrt{|h|}f(\phi)K^- +\int_{\Sigma} \D^{n-1}y \sqrt{|h|}{\cal L}_{\Sigma}^{(m)},\label{action-Sigma-J-app}
\end{align}
where $\Sigma_{\pm}$ are the sides of $\Sigma$ with normal vectors $n^{\mu}_{\pm}$ pointing outward, so that the boundary of each ${\cal M_{\pm}}$ is $\partial {\cal M_{\pm}}=(\partial{\cal M}_{\pm}-\Sigma_{\pm} ) \cup \Sigma_{\pm} $, and $\epsilon_+$, $\epsilon_-$, $\varepsilon$ independently take their values $\pm 1$ and $\phi^\pm|_{\Sigma}=\phi$. 
Here we used $z_{\pm}^i$ and $\zeta^{\pm}_{ij}$ for the coordinates and induced metric on the boundaries $\partial{\cal M}_{\pm}-\Sigma_{\pm}$, respectively.

From the results obtained in the previous subsections, variation of the action (\ref{action-Sigma-J-app}) with the boundary conditions $\delta g^{\pm\mu\nu}=\delta\phi^\pm=0$ 
at $\partial{\cal M}_{\pm}-\Sigma_{\pm}$ leads 
\begin{align} \label{delta_g I_J}
\delta_g I_{\rm J}=&\int_{\cal M_+} \D^{n}x_+\sqrt{-g^+}\biggl({\cal E}^+_{\mu\nu}-\frac12T^+_{\mu\nu}\biggl)\delta g^{+\mu\nu}+\int_{\cal M_-} \D^{n}x_-\sqrt{-g^-}\biggl({\cal E}^-_{\mu\nu}-\frac12T^-_{\mu\nu}\biggl)\delta g^{-\mu\nu} \nonumber \\
&-\int_{\Sigma_{+}} \D^{n-1}y \sqrt{|h|}{\cal F}^+_{\mu\nu}\delta g^{\mu\nu}
-\int_{\Sigma_{-}} \D^{n-1}y \sqrt{|h|}{\cal F}^-_{\mu\nu}\delta g^{\mu\nu}
+\int_{\Sigma} \D^{n-1}y \sqrt{|h|}\biggl(-\frac12t_{\mu\nu}\biggl)\delta g^{\mu\nu}  \nonumber \\
&
+(\mbox{Surface integral at $\partial\partial {\cal M^{\pm}}$}),
\end{align}
where
\begin{equation}
t_{\mu\nu}:=-2\frac{\partial {\cal L}^{(m)}_{\Sigma}}{\partial g^{\mu\nu}}+g_{\mu\nu}{\cal L}^{(m)}_{\Sigma},
\end{equation}
and 
\begin{align}
\delta_\phi I_{\rm J}=&\int_{\cal M_+}\D^{n}x_+ \sqrt{-g^+}{\cal E}^+_{(\phi)}\delta\phi^++\int_{\cal M_-}\D^{n}x_- \sqrt{-g^-}{\cal E}^-_{(\phi)}\delta\phi^- \nonumber \\
&+\int_{\Sigma_+}\D^{n-1}y\sqrt{|h|}{\cal F}^+_{(\phi)}\delta\phi+\int_{\Sigma_-}\D^{n-1}y\sqrt{|h|}{\cal F}^-_{(\phi)}\delta\phi. \label{delta_phi I_J}
\end{align} 

Choosing the normal vector $n^{\mu}$ to $\Sigma$ such that it points from $M_-$ to $M_+$, we have $n^{\mu}_-=n^{\mu}=-n^{\mu}_+$ and hence $K^+(n^{\mu}_+)=-K^+(n^{\mu})$ and $K^-(n^{\mu}_-)=K^-(n^{\mu})$.
They show ${ \cal F}^+_{\mu\nu}(n^{\mu}_+)=-{ \cal F}^+_{\mu\nu}(n^{\mu})$, 
${\cal F}^-_{\mu\nu}(n^{\mu}_-)={ \cal F}^-_{\mu\nu}(n^{\mu})$, ${\cal F}^+_{(\phi)}(n^{\mu}_+)=-{\cal F}^+_{(\phi)}(n^{\mu})$, and ${\cal F}^-_{(\phi)}(n^{\mu}_+)={\cal F}^-_{(\phi)}(n^{\mu})$, and consequently we have
\begin{align}
\int_{\Sigma_{+}} \D^{n-1}y \sqrt{|h|}{\cal F}^+_{\mu\nu}\delta g^{\mu\nu}=&-\int_{\Sigma} \D^{n-1}y \sqrt{|h|}{\cal F}^+_{\mu\nu}\delta g^{\mu\nu}, \\
\int_{\Sigma_{-}} \D^{n-1}y \sqrt{|h|}{\cal F}^-_{\mu\nu}\delta g^{\mu\nu}=&\int_{\Sigma} \D^{n-1}y \sqrt{|h|}{\cal F}^-_{\mu\nu}\delta g^{\mu\nu},\\
\int_{\Sigma_+}\D^{n-1}y\sqrt{|h|}{\cal F}^+_{(\phi)}\delta\phi=&-\int_{\Sigma}\D^{n-1}y\sqrt{|h|}{\cal F}^+_{(\phi)}\delta\phi, \\
\int_{\Sigma_-}\D^{n-1}y\sqrt{|h|}{\cal F}^-_{(\phi)}\delta\phi=&\int_{\Sigma}\D^{n-1}y\sqrt{|h|}{\cal F}^-_{(\phi)}\delta\phi.
\end{align}
Therefore, Eqs.~(\ref{delta_g I_J}) and (\ref{delta_phi I_J}) finally reduce to
\begin{align} \label{delta_g I_J2}
\delta_g I_{\rm J}=&\int_{\cal M_+} \D^{n}x_+\sqrt{-g^+}\biggl({\cal E}^+_{\mu\nu}-\frac12T^+_{\mu\nu}\biggl)\delta g^{+\mu\nu}+\int_{\cal M_-} \D^{n}x_-\sqrt{-g^-}\biggl({\cal E}^-_{\mu\nu}-\frac12T^-_{\mu\nu}\biggl)\delta g^{-\mu\nu} \nonumber \\
&+\int_{\Sigma} \D^{n-1}y \sqrt{|h|}\biggl({\cal F}^+_{\mu\nu}-{\cal F}^-_{\mu\nu} -\frac12t_{\mu\nu}\biggl)\delta g^{\mu\nu} 
+(\mbox{Surface integral at $\partial\partial {\cal M}^{\pm}$}),
\end{align}
and
\begin{align}
\delta_\phi I_{\rm J}=&\int_{\cal M_+}\D^{n}x_+ \sqrt{-g^+}{\cal E}^+_{(\phi)}\delta\phi^++\int_{\cal M_-}\D^{n}x_- \sqrt{-g^-}{\cal E}^-_{(\phi)}\delta\phi^- \nonumber \\
&-\int_{\Sigma}\D^{n-1}y\sqrt{|h|}({\cal F}^+_{(\phi)}-{\cal F}^-_{(\phi)})\delta\phi. \label{delta_phi I_J2}
\end{align} 

Hence, by the variational principle, we obtain the Einstein equations ${\cal E}^\pm_{\mu\nu}=(1/2)T^\pm_{\mu\nu}$ and the equation of motion for a scalar field ${\cal E}^\pm_{(\phi)}=0$ in the bulk spacetimes ${\cal M_\pm}$ as well as the junction conditions $[{\cal F}_{\mu\nu}]=(1/2)t_{\mu\nu}$ and $[{\cal F}_{(\phi)}]=0$ on $\Sigma$.
The bulk field equations are in the following form:
\begin{align} 
&2f(\phi)G_{\mu\nu}+g_{\mu\nu}\biggl(\frac{1}{2}(\nabla\phi)^2+V(\phi) \biggl) \nonumber \\
&~~~~~~~-(\nabla_\mu \phi)(\nabla_\nu \phi) -2\nabla_\mu\nabla_\nu f(\phi) +2g_{\mu\nu}\Box f(\phi)=T_{\mu\nu},\\
&\Box\phi +f'(\phi)R-V'(\phi)=0,
\end{align}
where we have omitted $\pm$ sign for simplicity, while the junction conditions at $\Sigma$ are written as
\begin{align}  
&2\varepsilon f(\phi)\left([K] h_{\mu\nu}-[K_{\mu\nu}]\right) +2\varepsilon f'(\phi) n^\sigma[\nabla_\sigma \phi]h_{\mu\nu}=t_{\mu\nu},\label{jc_other}\\
&2f'(\phi)[K]-n^\mu[\nabla_\mu\phi]=0. \label{jc_otherr}
\end{align}


\begin{thebibliography}{99}

\bibitem{Israel:1966rt}
W.~Israel,
Nuovo Cim.\ B {\bf 44S10} (1966) 1
[Nuovo Cim.\ B {\bf 44} (1966) 1]
Erratum: [Nuovo Cim.\ B {\bf 48} (1967) 463].

\bibitem{vacuumshell}
E.~Gravanis and S.~Willison,
Phys. Rev. D {\bf 75}, 084025 (2007);
C.~Garraffo, G.~Giribet, E.~Gravanis, and S.~Willison,
J. Math. Phys. {\bf 49}, 042502 (2008);
M.A.~Ram\'{\i}rez,
Class. Quant. Grav. {\bf 35}, 085004 (2018).



\bibitem{Poissonbook}
E. Poisson, A Relativist's Toolkit (Cambridge University Press, 2004).


\bibitem{Barrabes:1991ng}
C.~Barrab\`{e}s and W.~Israel,
Phys.\ Rev.\ D {\bf 43} (1991) 1129.


\bibitem{Barrabes:1995nk}
C.~Barrab\`{e}s and V.~P.~Frolov,
Phys.\ Rev.\ D {\bf 53} (1996) 3215.

\bibitem{Barrabes:1997kh}
C.~Barrab\`{e}s, G.~F.~Bressange and P.~A.~Hogan,
Phys.\ Rev.\ D {\bf 55} (1997) 3477.


\bibitem{Barrabes:1998rp}
C.~Barrab\`{e}s and P.~A.~Hogan,
Phys.\ Rev.\ D {\bf 58} (1998) 044013.
\bibitem{Barrabes:2000hm}
C.~Barrab\`{e}s and P.~A.~Hogan,
Class.\ Quant.\ Grav.\  {\bf 17} (2000) 4667.
\bibitem{Barrabes:2000es}
C.~Barrab\`{e}s and P.~A.~Hogan,
Phys.\ Lett.\ A {\bf 272} (2000) 226.
\bibitem{Barrabes:2001vy}
C.~Barrab\`{e}s and P.~A.~Hogan,
Int.\ J.\ Mod.\ Phys.\ D {\bf 10} (2001) 711.

\bibitem{Poisson:2002nv}
E.~Poisson,
``A Reformulation of the Barrab\`{e}s-Israel null shell formalism'',
gr-qc/0207101.


 	
\bibitem{j-conditions}
M.~Mars and J.M.M.~Senovilla,
Class. Quant. Grav. {\bf 10}, 1865 (1993);
J.M.M.~Senovilla,
Phys. Rev. D {\bf 88}, 064015 (2013);
J.M.M.~Senovilla,
Class. Quant. Grav. {\bf 31}, 072002 (2014);
B.~Reina, J.M.M.~Senovilla, R.~Vera,
Class. Quant. Grav. {\bf 33}, 105008 (2016);
J.M.M.~Senovilla,
JHEP {\bf 1811}, 134 (2018).

 	




 	
\bibitem{cr1999} 
H.~A.~Chamblin and H.~S.~Reall,
Nucl. Phys. {\bf B562}, 133 (1999).


  
\bibitem{Parattu:2015gga} 
  K.~Parattu, S.~Chakraborty, B.~R.~Majhi and T.~Padmanabhan,
  Gen.\ Rel.\ Grav.\  {\bf 48}, no. 7, 94 (2016).
 
\bibitem{Lehner:2016vdi} 
  L.~Lehner, R.~C.~Myers, E.~Poisson and R.~D.~Sorkin,
  Phys.\ Rev.\ D {\bf 94}, 084046 (2016).
 
 
\bibitem{Chakraborty:2016yna} 
  S.~Chakraborty,
  Fundam.\ Theor.\ Phys.\  {\bf 187}, 43 (2017).

 
\bibitem{Padilla:2012ze} 
  A.~Padilla and V.~Sivanesan,
  JHEP {\bf 1208}, 122 (2012).
  


\bibitem{Sakai:1992ud}
N.~Sakai and K.-i.~Maeda,
Prog.\ Theor.\ Phys.\  {\bf 90} (1993) 1001.

  \bibitem{Nishi:2014bsa} 
  S.~Nishi, T.~Kobayashi, N.~Tanahashi and M.~Yamaguchi,
  JCAP {\bf 1403}, 008 (2014).


\bibitem{Barcelo:2000js}
C.~Barcel\'{o} and M.~Visser,
Phys.\ Rev.\ D {\bf 63} (2001) 024004.


\bibitem{Barrabes:1997kk}
C.~Barrab\`{e}s and G.~F.~Bressange,
Class.\ Quant.\ Grav.\  {\bf 14} (1997) 805.


\bibitem{Bressange:1997ey}
G.~F.~Bressange,
Class.\ Quant.\ Grav.\  {\bf 15} (1998) 225.





\bibitem{wald}
R.~M.~Wald, {\it General Relativity},
(University of Chicago Press, 1984).

\bibitem{fn2007}
V.~Faraoni and S.~Nadeau,
Phys. Rev. D {\bf 75}, 023501 (2007).


\bibitem{roberts2014}
M.D.~Roberts,
Phys. Lett. {\bf B795}, 327 (2019).

\bibitem{maeda2015}
H.~Maeda,
Class. Quant. Grav. {\bf 32}, 135025 (2015).
\bibitem{roberts1989}
M.D.~Roberts, 
Gen. Rel. Grav.
{\bf 21}, 907 (1989). 
\bibitem{maeda2009}
H.~Maeda,
Phys. Rev. D {\bf 79}, 024030 (2009).

\bibitem{Aviles:2018vnf}
L.~Avil\'{e}s, H.~Maeda and C.~Mart\'{\i}nez,
Class. Quant. Grav. {\bf 35}, 245001 (2018).


\bibitem{Xu:2014xqa} 
W.~Xu,
Phys.\ Lett.\ B {\bf 738}, 472 (2014).


\bibitem{Gibbons-Hawking} 
G.~Gibbons and S.W.~Hawking, Phys. Rev. D {\bf 15}, 2752 (1977).
\bibitem{davis2003}
S.C.~Davis,
Phys. Rev. D {\bf 67}, 024030 (2003).
\bibitem{gw2003}
E.~Gravanis and S.~Willison, 
Phys. Lett. {\bf B562}, 118 (2003). 
\bibitem{mo2007}
O.~Mi\v{s}kovi\'c and R.~Olea,
JHEP {\bf 0710}, 028 (2007).

\bibitem{Lovelock} 
D.~Lovelock, 
J. Math. Phys. {\bf 12}, 498 (1971);
J. Math. Phys. {\bf 13}, 874 (1972).




\end{thebibliography}
\end{document}